\documentclass[a4paper, final, 12pt]{article}
\usepackage{amsmath, amssymb, latexsym, amscd, amsthm,amsfonts,amstext}
\usepackage[mathscr]{eucal}
\usepackage{graphicx}
\usepackage{subfig}
\usepackage{float}
\usepackage{color}
\usepackage{hyperref}
\usepackage[utf8]{inputenc}
\usepackage[english]{babel}

\numberwithin{equation}{section}
\input{tcilatex}
\begin{document}

\title{A New Type of Ill-Posed and Inverse Problems for Parabolic Equations}
\author{ Michael V. Klibanov \thanks{
Department of Mathematics and Statistics, University of North Carolina at
Charlotte, Charlotte, NC, 28223, USA, mklibanv@charlotte.edu.}}
\date{}
\maketitle

\begin{abstract}
The time dependent experimental data are always collected at discrete grids
with respect to the time t. The step size $h$ of such a grid is always
separated from zero by a certain positive number. The same is true for all
computations, which are always done on discrete grids with their grid step
sizes being not too small. These applied considerations prompt us to
introduce a new type of Ill-Posed Problems and Coefficient Inverse Problems
(CIP) for parabolic equations. In these problems the t-derivatives of
corresponding parabolic operators are written in finite differences with the
grid step size being separated from zero. We call this the \textquotedblleft
t-finite difference framework" (TFD). We address three long standing open
questions within the TFD framework. Finally, a numerical method is developed
for the CIP of monitoring epidemics. The global convergence of this method
is proven.
\end{abstract}



\section{Introduction}

\label{sec:1}

The time dependent experimental data are always collected at discrete grids
with respect to the time $t$. The step size $h$ of such a grid is always
separated from zero by a certain positive number. The same is true for all
computations, which are always done on discrete grids with their grid step
sizes being separated from zero. In other words, in applications one always
has 
\begin{equation}
h\geq h_{0}>0,  \label{1.1}
\end{equation}%
where the minimal possible grid step size $h_{0}$ is fixed. Therefore, it
makes sense, from the applied standpoint, to introduce a new type of both
Ill-Posed Problems and Coefficient Inverse Problems (CIPs) for parabolic
Partial Differential Equations (PDEs). In these problems, the $t-$%
derivatives of the parabolic operators are written in the form of finite
differences with condition (\ref{1.1}). Respectively, integrations with
respect to $t$ are done in the discrete sense. Note that all CIPs are
nonlinear ones.

We call the above \textquotedblleft $t-$finite difference framework" (TFD).
The goal of the current paper is to introduce this new class of problems for
parabolic equations and to investigate some of its properties. \ We
demonstrate that three long standing open questions can be addressed within
the TFD framework. These questions are listed below in this section.

The TFD framework reduces the problems we consider here to coupled systems
of elliptic PDEs. Each of these PDEs corresponds to a grid point. Each
equation of that system has both Dirichlet and Neumann boundary conditions
on at least a part of the boundary. Next, applications of Carleman estimates
to this system leads to the target results.

Let $T>0$ be a number and $\Omega \subset \mathbb{R}^{n}$ be a bounded
domain with a piecewise smooth boundary $\partial \Omega .$ Let $\Gamma
\subseteq \partial \Omega $ be a part of this boundary, and let $\Gamma \in
C^{3}$. Below $x\in \Omega $ is the spatial variable. Denote%
\begin{equation*}
\Omega _{T}=\Omega \times \left( 0,T\right) ,\text{ }S_{T}=\partial \Omega
\times \left( 0,T\right) ,\text{ }\Gamma _{T}=\Gamma \times \left(
0,T\right) .
\end{equation*}%
\ Hence, $\Omega _{T}$ is the time cylinder, $S_{T}$ is its lateral boundary
and $\Gamma _{T}$ is a part of the lateral boundary.

Suppose that both Dirichlet and Neumann conditions for a solution of a
parabolic equation are given at $\Gamma _{T}.$ Then these conditions are
called \textquotedblleft lateral Cauchy data". The problem of finding the
solution of that equation using lateral Cauchy data is well known under the
name \textquotedblleft Unique Continuation Problem" (UCP), see, e.g. \cite%
{Isakov,KL,LRS}. The problem of finding an unknown $x-$dependent coefficient
of that equation using the initial data as well as the lateral Cauchy data
is called Coefficient Inverse Problem (CIP). It turns out that the method of 
\cite{BukhKlib} reduces such a CIP to a UCP for a nonlinear integral
differential equation, which contains the principal part of the originating
parabolic operator as well as Volterra-like integrals with respect to $t$.
These are two types of integrals: either 
\begin{equation}
\dint\limits_{t_{0}}^{t}\left( \cdot \right) d\tau ,\text{ }t_{0}>0,\text{ }%
t\in \left( 0,T\right) ,  \label{1.2}
\end{equation}%
or%
\begin{equation}
\dint\limits_{0}^{t}\left( \cdot \right) d\tau ,\text{ }t\in \left(
0,T\right) .  \label{1.3}
\end{equation}%
In the case (\ref{1.2}) one assumes the knowledge of the solution of the
generating parabolic PDE at the moment of time $\left\{ t=t_{0}\right\} ,$
in addition to the lateral Cauchy data, and the knowledge of the initial
condition at $\left\{ t=0\right\} $ is not assumed. In this case the method
of \cite{BukhKlib} enables one to prove H\"{o}lder stability estimates and
uniqueness for corresponding CIPs, see, e.g. \cite%
{Isakov,Klib841,Klib842,Ksurvey,KL}, and even Lipschitz stability, see, e.g. 
\cite{Yam}. Therefore, we do not consider the case (\ref{1.2}) below.

Integrals (\ref{1.3}) take place in such CIPs for parabolic PDEs, in which
the lateral Cauchy data are complemented with the initial condition at $%
\left\{ t=0\right\} ,$ which is more natural than the case of the data at $%
\left\{ t=t_{0}\right\} .$ \underline{Nothing is known about such CIPs.}%
\emph{\ }On the other hand, the TFD framework enables us to handle this case.

We address in this paper three long standing open questions within the TFD
framework:

\begin{enumerate}
\item The problem of H\"{o}lder and Lipschitz stability estimates for
nonlinear UCPs for parabolic PDEs in the entire time interval $t\in \left(
0,T\right) .$ On the other hand, only estimates in the interval $t\in \left(
\varepsilon ,T-\varepsilon \right) ,\varepsilon \in \left( 0,T/2\right) $
are known now both in linear and nonlinear cases. The only two exceptions
are stability estimates of \cite{Bourg,Klib2006}. However, they are
logarithmic ones, which means that they weaker than H\"{o}lder stability
estimates.

\item The problem of H\"{o}lder and Lipschitz stability estimates in the
entire time interval $t\in \left( 0,T\right) $ for nonlinear UCPs for the
above mentioned integral differential equations with Volterra integrals (\ref%
{1.3}).

\item Results of item 2 imply H\"{o}lder and Lipschitz stability estimates \
for a broad class of CIPs for parabolic PDEs in the case when the lateral
Cauchy data are complemented with the initial condition at $\left\{
t=0\right\} $
\end{enumerate}

Finally, we demonstrate the applicability of the TFD framework to a specific
applied example. More precisely, we use the TFD framework to prove the
Lipschitz stability estimate for a CIP of monitoring epidemics using
boundary measurements. A system of three nonlinear coupled parabolic
equations is involved. First, we prove Lipschitz stability estimate for this
CIP. Next, we construct a globally convergent numerical method for this
problem and provide a convergence analysis for it. This is a version of the
so-called convexification method. Another version of the convexification for
this problem was developed in \cite{epid}, where the lateral Cauchy data
were combined with the data at $\left\{ t=t_{0}>0\right\} $ with resulting
Volterra integrals (\ref{1.2}) instead of the more challenging case (\ref%
{1.3}) which we work here within the TFD framework. Stability estimates were
not obtained in \cite{epid}.

The development of the convexification method is prompted by our desire to
avoid the well known phenomenon of multiple local minima and ravines of
conventional least squares cost functionals for CIPs \cite{B,Gonch,Riz}.
More precisely, the convexification replaces the local convergence of the
conventional numerical methods for CIPs with the global convergence, see
Remark 9.1 about the global convergence property. The development of the
convexification was originated in \cite{Klib97} and continued since then,
see, e.g. \cite{Bak,Baud1,Baud2,Baud3,Kpar,KL,SAR,epid} for some references.
A notable feature of the convexification method of this paper, which was not
the case of other above cited references, is that the penalty regularization
term is not involved here.

All functions considered below are real valued ones. In section 2 we
introduce the $t-$finite differences. In section 3 we introduce statements
of our UCP and the first CIP, which we call CIP1. In section 4 we present
the TFD framework for CIP1. In section 5 we formulate two Carleman estimates
and prove one of them. In section 6 we prove H\"{o}lder and Lipschitz
stability estimates for our UCP. In section 7 we formulate H\"{o}lder and
Lipschitz stability estimates for CIP1. Actually theorems of section 7
easily follow from theorems of section 6. In section 8 we first formulate
our CIP of monitoring of epidemics using boundary measurements. This is
CIP2. Next, we reformulate CIP2 in the TFD framework and formulate Lipschitz
stability estimate for so reformulated CIP2. That estimate easily follows
from one of theorems of section 6. In section 9, we formulate the
convexification method for the TFD framework for CIP2 and provide its
convergence analysis.

\section{$t-$Finite Differences Framework (TFD)}

\label{sec:2}

\subsection{Sets and spaces}

\label{sec:2.1}

Consider the partition of the interval $\left[ 0,T\right] $ in $k\geq 3$
subintervals,%
\begin{equation}
0=t_{0}<t_{1}<...<t_{k-1}<t_{k}=T,\text{ }t_{i}-t_{i-1}=h,\text{ }i=1,...,k.
\label{2.2}
\end{equation}%
Denote%
\begin{equation}
Y=\left\{ t_{i}\right\} _{i=0}^{k}.  \label{2.20}
\end{equation}%
Define semi-discrete analogs of sets (\ref{1.2}) are: 
\begin{equation}
\Omega _{h,T}=\Omega \times Y,\text{ }S_{h,T}=\partial \Omega \times Y,\text{
}\Gamma _{h,T}=\Gamma \times Y.  \label{2.21}
\end{equation}%
Similarly, for any set $\Psi \subset \mathbb{R}^{n}$%
\begin{equation}
\Psi _{h,T}=\Psi \times Y.  \label{2.210}
\end{equation}%
Therefore, any function defined on $\Psi _{h,T}$ is actually a $\left(
k+1\right) -$dimensional vector function, 
\begin{equation}
u\left( x,t\right) =\left( u\left( x,t_{0}\right) ,u\left( x,t_{1}\right)
,...,u\left( x,t_{k}\right) \right) ^{T}\text{ for }\left( x,t\right) \in
\Psi _{h,T}.  \label{2.3}
\end{equation}%
Let $m\in \left[ 0,3\right] $ be an integer. We now define some spaces
associated with vector functions (\ref{2.3}). Let $\Psi \subset \mathbb{R}%
^{n}$ be either a bounded domain or a hypersurface such that $\Psi \in
C^{3}. $ Then, using (\ref{2.210}), we set%
\begin{equation}
\left. 
\begin{array}{c}
C^{m}\left( \overline{\Psi }_{h,T}\right) =\left\{ u:\left\Vert u\right\Vert
_{C^{m}\left( \overline{\Psi }_{h,T}\right) }=\left(
\dsum\limits_{i=0}^{k}\left\Vert u\left( x,t_{i}\right) \right\Vert
_{C^{m}\left( \overline{\Psi }\right) }^{2}\right) ^{1/2}<\infty \right\} ,
\\ 
H^{m}\left( \Psi _{h,T}\right) =\left\{ u:\left\Vert u\right\Vert
_{H^{m}\left( \Psi _{h,T}\right) }=\left( \dsum\limits_{i=0}^{k}\left\Vert
u\left( x,t_{i}\right) \right\Vert _{H^{m}\left( \Psi \right) }^{2}\right)
^{1/2}<\infty \right\} .%
\end{array}%
\right.  \label{2.03}
\end{equation}%
In particular, $C^{0}\left( \overline{\Psi }_{h,T}\right) =C\left( \overline{%
\Psi }_{h,T}\right) $ and $H^{0}\left( \Psi _{h,T}\right) =L_{2}\left( \Psi
_{h,T}\right) .$ Recall that $\Gamma \subseteq \partial \Omega ,\Gamma \in
C^{3}.$ We set 
\begin{equation}
H^{m}\left( \Gamma _{h,T}\right) =\left\{ u:\left\Vert u\right\Vert
_{H^{m}\left( \Gamma _{h,T}\right) }=\left( \dsum\limits_{i=0}^{k}\left\Vert
u\left( x,t_{i}\right) \right\Vert _{H^{m}\left( \Gamma \right) }^{2}\right)
^{1/2}<\infty \right\} ,  \label{2.30}
\end{equation}%
We set%
\begin{equation}
\partial \Omega =\Gamma \cup _{j=1}^{s}\partial _{j}\Omega ,\text{ }\partial
_{j}\Omega \in C^{3},\text{ }\partial _{j}\Omega _{h,T}=\partial _{j}\Omega
\times Y.  \label{2.31}
\end{equation}%
Using (\ref{2.21}) and (\ref{2.210}), we set similarly with (\ref{2.30})%
\begin{equation}
H^{m}\left( S_{h,T}\right) =\left\{ 
\begin{array}{c}
u:\left\Vert u\right\Vert _{H^{m}\left( S_{h,T}\right) }= \\ 
=\left( \left\Vert u\right\Vert _{H^{m}\left( \Gamma _{h,T}\right)
}^{2}+\dsum\limits_{j=1}^{s}\dsum\limits_{i=0}^{k}\left\Vert u\left(
x,t_{i}\right) \right\Vert _{H^{m}\left( \text{ }\partial _{j}\Omega
_{h,T}\right) }^{2}\right) ^{1/2}<\infty%
\end{array}%
\right\} .  \label{2.310}
\end{equation}

\textbf{Remarks 2.1. }

\emph{1. It is well known that when discussing questions of existence and
uniqueness of initial boundary value problems for parabolic equations,
spaces like }$C^{2r,r},C^{2r+\rho ,r+\rho /2},H^{2r,r}$\emph{\ are
conventionally used, where }$r\geq 1$\emph{\ is an integer and }$\rho \in
\left( 0,1\right) ,$\emph{\ see, e.g. \cite{Ladpar}. However, since we do
not discuss these questions here and also since we are interested in
stability estimates for our ill-posed problems within the TFD framework,
then we will use} \emph{spaces }$C^{r}$\emph{\ in the continuos cases and
spaces (\ref{2.03}), (\ref{2.30}), (\ref{2.310}) in the semi-discrete cases.
Furthermore, even though in some of our stability estimates we will use }$%
H^{2}\left( \Omega _{h,T}\right) -$\emph{norms\ of solutions of our
resulting BVPs}$,$\emph{\ we will still assume that the lateral Cauchy data
belong to the spaces }$H^{2}\left( \Gamma _{h,T}\right) ,H^{2}\left(
S_{h,T}\right) .$\emph{\ By trace theorem, the latter can be guaranteed if
requiring the }$H^{3}\left( \Omega _{h,T}\right) -$\emph{smoothness of those
solutions. }

2. \emph{It is well known that minimal smoothness assumptions are minor
concerns in the field of CIPs, see, e.g. \cite{Nov} and \cite[Theorem 4.1]%
{Rom}. Therefore, they are also minor concerns here.}

\subsection{$t-$finite differences and discrete Volterra integrals}

\label{sec:2.2}

We now define $t-$finite differences $f_{h}^{\prime }\left( t\right) $ of
the function $f\left( t\right) \in C^{3}\left[ 0,T\right] $ on the discrete
set (\ref{2.2}). We set%
\begin{equation}
\partial _{h,t}f\left( t_{i}\right) =\frac{f\left( t_{i+1}\right) -f\left(
t_{i-1}\right) }{2h},\text{ }i=1,...,k-1.  \label{2.4}
\end{equation}%
Expression (\ref{2.4}) is valid for interior points $t_{i}$ of the grid (\ref%
{2.2}). Temporary allowing $h\rightarrow 0^{+},$ we note that it is well
known that the second order approximation accuracy $O\left( h^{2}\right) $
is delivered by (\ref{2.4}). In addition, we also need the finite difference
derivatives at the edge points $t_{0}=0$ and $t_{k}=T.$ It is easy to verify
that the following two formulas deliver the second order accuracy at $%
t_{0}=0 $ and at $t_{k}=T:$%
\begin{equation}
\left. \partial _{h,t}f\left( t_{0}\right) =\frac{3f\left( t_{0}\right)
-4f\left( t_{1}\right) +f\left( t_{2}\right) }{2h},\right.  \label{2.5}
\end{equation}%
\begin{equation}
\partial _{h,t}f\left( t_{k}\right) =\frac{3f\left( t_{k}\right) -4f\left(
t_{k-1}\right) +f\left( t_{k-2}\right) }{2h}.  \label{2.6}
\end{equation}

\textbf{Remark 2.2.} \emph{It is an important\ observation for our method
that values }$f\left( t_{i}\right) $\emph{\ at all points }$t_{i}\in Y$ 
\emph{are involved in the right hand sides of (\ref{2.4})-(\ref{2.6}).}

We also define the discrete Volterra integral of the function $f\left(
t\right) $ via the trapezoidal rule, which gives the $O\left( h^{2}\right) $
accuracy,%
\begin{equation}
\left. \left( \dint\limits_{0}^{t_{i}}fd\tau \right) _{h}=\frac{h}{2}%
\dsum\limits_{j=1}^{i}\left( f\left( t_{j-1}\right) +f\left( t_{j}\right)
\right) ,\text{ }t_{i}\in Y,\text{ }i=1,...,k,\right.  \label{2.7}
\end{equation}%
\begin{equation}
\left( \dint\limits_{0}^{t_{i}}fd\tau \right) _{h}=0,\text{ }i=0.
\label{2.8}
\end{equation}%
Thus, formulas (\ref{2.7}), (\ref{2.8}) define discrete Volterra integrals
for all $t_{i}\in Y,$ where the set $Y$ is defined in (\ref{2.20}).

\section{Statements of Problems}

\label{sec:3}

\subsection{Unique continuation problem for a nonlinear integral
differential equation}

\label{sec:3.1}

We now formulate the unique continuation problem for a nonlinear integral
differential equation in the TFD framework. Let the vector function $u\in
C^{2}\left( \overline{\Omega }_{h,T}\right) .$ Introduce the vector function 
$P\left( u,x,t_{s}\right) ,$%
\begin{equation}
\left. 
\begin{array}{c}
P\left( u,x,t_{s}\right) = \\ 
=\left( \left( \dint\limits_{0}^{t_{s}}\nabla _{x}u\left( x,\tau \right)
d\tau d\tau \right) _{h},\left( \dint\limits_{0}^{t_{s}}u\left( x,\tau
\right) d\tau \right) _{h},\left( \dint\limits_{0}^{t_{s}}\partial
_{h,t}u\left( x,\tau \right) d\tau \right) _{h}\right) , \\ 
x\in \Omega ,\text{ }t_{s}\in Y,s=0,...,k.%
\end{array}%
\right.  \label{3.0}
\end{equation}%
The nonlinear integral differential equation with discrete Volterra
integrals (\ref{2.7}), (\ref{2.8}) in it is : 
\begin{equation}
\left. 
\begin{array}{c}
\partial _{h,t}u\left( x,t_{s}\right) = \\ 
=F\left( u_{x_{i}x_{j}},\nabla _{x}u,u,P\left( u,x,t_{s}\right)
,x,t_{s}\right) , \\ 
i,j=1,...,n;\text{ }x\in \Omega ,\text{ }t_{s}\in Y,%
\end{array}%
\right.  \label{3.1}
\end{equation}%
where the function $F\left( y,x,t_{s}\right) \in C^{2}\left( \mathbb{R}%
^{N}\times \overline{\Omega }_{h,T}\right) ,$ where $N$ is the total number
of arguments in the vector $y$, which includes all terms with all components
of the vector function $u\in C^{2}\left( \overline{\Omega }_{h,T}\right) ,$
its derivatives and discrete integrals in (\ref{3.0}), (\ref{3.1}).

Let $\mu >0$ be a number. We impose the following conditions on the function 
$F$:%
\begin{equation}
\frac{\partial F\left( y,x,t_{s}\right) }{\partial u_{x_{i}x_{j}}\left(
x,t_{s}\right) }=\frac{\partial F\left( y,x,t_{s}\right) }{\partial
u_{x_{j}x_{i}}\left( x,t_{s}\right) },\forall y\in \mathbb{R}^{N},\text{ }%
\forall \left( x,t_{s}\right) \in \Omega _{h,T},\forall i,j=1,...,n,
\label{3.2}
\end{equation}%
\begin{equation}
\mu \left\vert \xi \right\vert ^{2}\leq \dsum\limits_{i,j=1}^{n}\frac{%
\partial F\left( y,x,t_{s}\right) }{\partial u_{x_{i}x_{j}}\left(
x,t_{s}\right) }\xi _{i}\xi _{j},\text{ }\forall \xi \in \mathbb{R}^{n},%
\text{ }\forall y\in \mathbb{R}^{N},\forall \left( x,t_{s}\right) \in \Omega
_{h,T}.  \label{3.3}
\end{equation}%
In the specific case of (\ref{3.2}), (\ref{3.3}), $u_{x_{i}x_{j}}\left(
x,t_{s}\right) $ means the coordinate of the vector $y\in \mathbb{R}^{N},$
which corresponds to $u_{x_{i}x_{j}}\left( x,t_{s}\right) $ in (\ref{3.1}).
Obviously conditions \ref{3.2}), (\ref{3.3}) are analogs of the ellipticity
conditions. Hence, we call the operator $F$ in (\ref{3.1}) \textquotedblleft
nonlinear elliptic operator in $\Omega _{h,T}".$ Problem (\ref{3.0})-(\ref%
{3.4}) is exactly the Unique Continuation Problem. We are interested in
stability estimates for this problem.

\textbf{Stability Estimates for UCP (\ref{3.0})-(\ref{3.4}).} \emph{Given
notation (\ref{3.0}), suppose that two vector functions }$u_{1}\left(
x,t_{s}\right) ,u_{2}\left( x,t_{s}\right) \in C^{2}\left( \overline{\Omega }%
_{h,T}\right) $\emph{\ (see Remarks 2.1) are solutions of equation (\ref{3.1}%
) with the following lateral Cauchy data at }$\Gamma _{h,T}:$\emph{\ }%
\begin{equation}
u_{i}\mid _{\Gamma _{h,T}}=g_{i,0}\left( x,t_{s}\right) ,\text{ }\partial
_{l}u_{i}\mid _{\Gamma _{h,T}}=g_{i,1}\left( x,t_{s}\right) ,\text{ }%
t_{s}\in Y,\text{ }i=1,2,  \label{3.4}
\end{equation}%
\emph{where }$l$\emph{\ is the outward looking unit normal vector at }$%
\Gamma .$\emph{\ Assume that conditions (\ref{3.2}) and (\ref{3.3}) hold.
Estimate an appropriate norm \ of the difference} $u_{1}-u_{2}$\emph{\ via
certain norms of differences }$g_{1,0}-g_{2,0}$\emph{\ and }$%
g_{1,1}-g_{2,1}. $

\subsection{A Coefficient Inverse Problem}

\label{sec:3.2}

Let $\alpha =\left( \alpha _{1},...,\alpha _{n}\right) $ be a multii-index
with non-negative integer coordinates. Denote $\left\vert \alpha \right\vert
=\alpha _{1}+...+\alpha _{n}.$ Let $D_{x}^{\alpha }=D_{x_{1}}^{\alpha
_{1}}D_{x_{2}}^{\alpha _{2}}...D_{x_{n}}^{\alpha _{n}}$ be the differential
operator. For any vector $\xi =\left( \xi _{1},...,\xi _{n}\right) ^{T}\in 
\mathbb{R}^{n}$ set $\xi ^{\alpha }=\xi _{1}^{\alpha _{1}}...\xi
_{n}^{\alpha _{n}}.$ Consider functions $a_{\alpha }\left( x\right) $ for $%
\left\vert \alpha \right\vert =2$ and $a_{\alpha }\left( x,t\right) $ for $%
\left\vert \alpha \right\vert \leq 1$ satisfying the following conditions%
\begin{equation}
\left. 
\begin{array}{c}
a_{ij}\left( x\right) =a_{ji}\left( x\right) ,\text{ }i,j=1,...,n;\text{ }%
x\in \Omega , \\ 
\mu \left\vert \xi \right\vert ^{2}\leq \dsum\limits_{\left\vert \alpha
\right\vert =2}^{n}a_{ij}\left( x\right) \xi _{i}\xi _{j},\text{ }\forall
x\in \overline{\Omega },\text{ }\forall \xi \in \mathbb{R}^{n}, \\ 
a_{\alpha }\in C^{1}\left( \overline{\Omega }\right) \text{ for }\left\vert
\alpha \right\vert =2,\text{ }\left\Vert a_{\alpha }\right\Vert
_{C^{1}\left( \overline{\Omega }\right) }\leq a^{0},%
\end{array}%
\right.  \label{3.5}
\end{equation}%
\begin{equation}
\left. 
\begin{array}{c}
a_{\alpha },\partial _{t}a_{\alpha }\in C\left( \overline{\Omega }%
_{T}\right) \text{ for }\left\vert \alpha \right\vert \leq 1, \\ 
\left\Vert a_{\alpha }\right\Vert _{C\left( \overline{\Omega }_{T}\right)
}\leq D,\text{ }\left\Vert \partial _{t}a_{\alpha }\right\Vert _{C\left( 
\overline{\Omega }_{T}\right) }\leq D\text{ for }\left\vert \alpha
\right\vert \leq 1.%
\end{array}%
\right.  \label{3.50}
\end{equation}%
Here $a^{0}>0$ and $D>0$ are certain numbers and $a_{ij}\left( x\right) $
are just different notations of $a_{\alpha }\left( x\right) $ with $%
\left\vert \alpha \right\vert =2.$ The first two lines of (\ref{3.5}) are
linear analogs of conditions (\ref{3.2}), (\ref{3.3}). Consider the linear
elliptic operator of the second order $L$ with its principal part $L_{0},$%
\begin{equation}
\left. 
\begin{array}{c}
Lu=\dsum\limits_{\left\vert \alpha \right\vert \leq 2}a_{\alpha }\left(
x,t\right) D_{x}^{\alpha }u=L_{0}u+L_{1}u, \\ 
L_{0}u=\dsum\limits_{\left\vert \alpha \right\vert =2}a_{\alpha }\left(
x\right) D_{x}^{\alpha }u,\text{ } \\ 
L_{1}u=\dsum\limits_{\left\vert \alpha \right\vert \leq 1}a_{\alpha }\left(
x,t\right) D_{x}^{\alpha }u.%
\end{array}%
\right.  \label{3.6}
\end{equation}%
Consider the parabolic equation with the initial condition at $\left\{
t=0\right\} $ and lateral Cauchy data at $\Gamma _{T},$%
\begin{equation}
u_{t}=Lu,\text{ }\left( x,t\right) \in \Omega _{T},  \label{3.7}
\end{equation}%
\begin{equation}
u\left( x,0\right) =f\left( x\right) ,  \label{3.8}
\end{equation}%
\begin{equation}
u\mid _{\Gamma _{T}}=p_{0}\left( x\right) ,\text{ }\partial _{l}u\mid
_{\Gamma _{T}}=p_{1}\left( x\right) .  \label{3.9}
\end{equation}

\textbf{Coefficient Inverse Problem 1 (CIP1).}\emph{\ Let coefficients of
the operator }$L$\emph{\ satisfy conditions (\ref{3.5}). Let }$\alpha _{0}$%
\emph{\ be a fixed multi-index, }$\left\vert \alpha _{0}\right\vert \leq 2.$%
\emph{\ Suppose that a coefficient }$a_{\alpha _{0}}=a_{\alpha _{0}}\left(
x\right) $\emph{\ of the operator }$L$\emph{\ in (\ref{3.6}) is unknown,
whereas other coefficients as well as right hand sides in (\ref{3.8}) and (%
\ref{3.9}) are known. Suppose that we have two pairs of functions }$\left(
u_{1},a_{1,\alpha _{0}}\right) $\emph{\ and }$\left( u_{2},a_{2,\alpha
_{0}}\right) $\emph{\ satisfying equation (\ref{3.7}) with the same initial
condition (\ref{3.8}), where functions }$u_{1},u_{2}\in C^{3}\left( 
\overline{\Omega }_{T}\right) $ \emph{(see Remarks 2.1)}$.$\emph{\ Assume
that }%
\begin{equation}
u_{i}\mid _{\Gamma _{T}}=p_{i,0}\left( x,t\right) ,\text{ }\partial
_{l}u_{i}\mid _{\Gamma _{T}}=p_{i,1}\left( x,t\right) ,\text{ }i=1,2.
\label{3.10}
\end{equation}%
\emph{Also, assume that }%
\begin{equation}
\left\vert D_{x}^{\alpha _{0}}f\left( x\right) \right\vert \geq c_{1}\text{
in }\Omega ,  \label{3.90}
\end{equation}%
\emph{where }$c_{1}>0$\emph{\ is a number. Estimate an appropriate norm \ of
the differences }$\widetilde{u}=u_{1}-u_{2},\widetilde{a}_{\alpha
_{0}}=a_{1,\alpha _{0}}-a_{2,\alpha _{0}}$\emph{\ via certain norms of
differences }$\widetilde{p}_{0}=p_{1,0}-p_{2,0}$\emph{\ and }$\widetilde{p}%
_{1}=p_{1,1}-p_{2,1}.$

\section{The TFD Framework for CIP1 (\protect\ref{3.7})-(\protect\ref{3.90})}

\label{sec:4}

For any two pairs of numbers $\left( b_{1},d_{1}\right) ,$ $\left(
b_{2},d_{2}\right) $%
\begin{equation}
\left. 
\begin{array}{c}
b_{1}d_{1}-b_{2}d_{2}=\widetilde{b}d_{1}+\widetilde{d}b_{2}, \\ 
\widetilde{b}=b_{1}-b_{2},\text{ }\widetilde{d}=d_{1}-d_{2}.%
\end{array}%
\right.  \label{4.1}
\end{equation}%
Hence, using (\ref{3.7})-(\ref{3.10}), (\ref{4.1}) and\emph{\ }notations of
the formulation of CIP1, we obtain%
\begin{equation}
\left. 
\begin{array}{c}
\widetilde{u}_{t}=L^{\left( 1\right) }\widetilde{u}+\widetilde{a}_{\alpha
_{0}}\left( x\right) D_{x}^{\alpha _{0}}u_{2},\text{ in }\Omega _{T}, \\ 
\widetilde{u}\left( x,0\right) =0, \\ 
\widetilde{u}\mid _{\Gamma _{T}}=\widetilde{p}_{0}\left( x,t\right) ,\text{ }%
\partial _{l}\widetilde{u}\mid _{\Gamma _{T}}=\widetilde{p}_{1}\left(
x,t\right) ,%
\end{array}%
\right.  \label{4.2}
\end{equation}%
where $L^{\left( 1\right) }$ is the operator $L$ in (\ref{3.6}), in which
the coefficient $a_{\alpha _{0}}\left( x\right) $ is replaced with $%
a_{1,\alpha _{0}}\left( x\right) .$ By (\ref{3.8}), (\ref{3.90}) and (\ref%
{4.2}) 
\begin{equation}
\widetilde{a}_{\alpha _{0}}\left( x\right) =\frac{\widetilde{u}_{t}\left(
x,0\right) }{D_{x}^{\alpha _{0}}f\left( x\right) }.  \label{4.3}
\end{equation}%
Denote 
\begin{equation}
w\left( x,t\right) =\widetilde{u}_{t}\left( x,t\right) .  \label{4.4}
\end{equation}%
Then $w\in H^{2}\left( \Omega _{T}\right) .$ The second line of (\ref{4.2})
implies: 
\begin{equation}
\widetilde{u}\left( x,t\right) =\dint\limits_{0}^{t}w\left( x,\tau \right)
d\tau .  \label{4.5}
\end{equation}%
Next, 
\begin{equation}
\widetilde{u}_{t}\left( x,0\right) =w\left( x,0\right) =w\left( x,t\right)
-\dint\limits_{0}^{t}w_{t}\left( x,\tau \right) d\tau .  \label{4.6}
\end{equation}%
Thus, (\ref{3.6}) and (\ref{4.2})-(\ref{4.6}) lead to the following BVP with
the lateral Cauchy data: 
\begin{equation*}
w_{t}=L^{\left( 1\right) }w+\dsum\limits_{\left\vert \alpha \right\vert \leq
1}\partial _{t}a_{\alpha }\left( x,t\right)
\dint\limits_{0}^{t}D_{x}^{\alpha }w\left( x,\tau \right) d\tau +
\end{equation*}%
\begin{equation}
+\frac{D_{x}^{\alpha _{0}}\partial _{t}u_{2}\left( x,t\right) }{%
D_{x}^{\alpha _{0}}f\left( x\right) }\left( w\left( x,t\right)
-\dint\limits_{0}^{t}w_{t}\left( x,\tau \right) d\tau \right) ,\text{ }%
\left( x,t\right) \in Q_{T},  \label{4.7}
\end{equation}%
\begin{equation}
w\mid _{\Gamma _{T}}=\partial _{t}\widetilde{p}_{0}\left( x,t\right) ,\text{ 
}\partial _{l}w\mid _{\Gamma _{T}}=\partial _{t}\widetilde{p}_{1}\left(
x,t\right) .  \label{4.8}
\end{equation}

Thus, we have transformed problem (\ref{4.2}) in BVP (\ref{4.7}), (\ref{4.8}%
), which does not contain the unknown function $\widetilde{a}_{\alpha
_{0}}\left( x\right) $ but has Volterra integrals. Likewise integral
differential equation (\ref{4.7}) does not have an initial condition at $%
\left\{ t=0\right\} .$ We are ready now to rewrite (\ref{4.7}), (\ref{4.8})
in the TFD framework.

\textbf{The TFD Framework for the Problem of Stability Estimates for CIP1:} 
\emph{Let condition (\ref{3.90}) be valid. Estimate the vector function }$%
v\in H^{2}\left( \Omega _{h,T}\right) $\emph{\ via vector functions }$%
\partial _{h,t}\widetilde{p}_{0}\left( x,t_{s}\right) $\emph{\ and }$%
\partial _{h,t}\widetilde{p}_{1}\left( x,t_{s}\right) ,$\emph{\ assuming
that the following conditions (\ref{4.9}), (\ref{4.10}) hold:}%
\begin{equation}
\left. 
\begin{array}{c}
\partial _{h,t}w\left( x,t_{s}\right) =\left( L^{\left( 1\right) }w\right)
\left( x,t_{s}\right) +\dsum\limits_{\left\vert \alpha \right\vert \leq
1}\left( \partial _{h,t}a_{\alpha }\left( x,t_{s}\right) \right) \left(
\dint\limits_{0}^{t_{s}}D_{x}^{\alpha }w\left( x,\tau \right) d\tau \right)
_{h}+ \\ 
+\left( D_{x}^{\alpha _{0}}\left( \partial _{h,t}u_{2}\left( x,t_{s}\right)
\right) \right) \left( D_{x}^{\alpha _{0}}f\left( x\right) \right) ^{-1} 
\left[ w\left( x,t_{s}\right) -\left( \dint\limits_{0}^{t_{s}}\partial
_{h,\tau }w\left( x,\tau \right) d\tau \right) _{h}\right] , \\ 
\text{ }x\in \Omega ,\text{ }t_{s}\in Y,%
\end{array}%
\right.  \label{4.9}
\end{equation}%
\begin{equation}
w\mid _{\Gamma _{h,T}}=\partial _{h,t}\widetilde{p}_{0}\left( x,t_{s}\right)
,\text{ }\partial _{l}w\mid _{\Gamma _{h,T}}=\partial _{h,t}\widetilde{p}%
_{1}\left( x,t_{s}\right) ,\text{ }t_{s}\in Y,  \label{4.10}
\end{equation}%
\emph{where Volterra integrals are understood as in (\ref{2.7}) and (\ref%
{2.8}), see Remark 2.2. Thus, BVP (\ref{4.9}), (\ref{4.10}) is a special
linear version of UCP (\ref{3.1})-(\ref{3.4}). In addition, estimate the
function }$\widetilde{a}_{\alpha _{0}}\left( x\right) $ 
\begin{equation}
\widetilde{a}_{\alpha _{0}}\left( x\right) =\frac{w\left( x,0\right) }{%
D_{x}^{\alpha _{0}}f\left( x\right) },\text{ }x\in \Omega .  \label{4.11}
\end{equation}

Formula (\ref{4.11}) is obtained from $t-$finite differences analogs of
formulas (\ref{4.3}) and (\ref{4.4}). The boundary data (\ref{4.10}) are
incomplete since one might have $\Gamma \neq \partial \Omega .$ Hence, we
will use (\ref{4.10}) for the H\"{o}lder stability estimate. However, for
the Lipschitz stability estimate we need complete data at $S_{h,T}.$ Thus,
along with (\ref{4.10}), we will also consider the following boundary data:%
\begin{equation}
w\mid _{S_{h,T}}=\partial _{h,t}\widetilde{p}_{0}\left( x,t_{s}\right) ,%
\text{ }\partial _{l}w\mid _{S_{h,T}}=\partial _{h,t}\widetilde{p}_{1}\left(
x,t_{s}\right) .  \label{4.12}
\end{equation}

\section{Two Carleman Estimates for Elliptic Operators}

\label{sec:5}

\subsection{Domains}

\label{sec:5.1}

Let $A_{1},A_{2}>0$ be two numbers. Denote $\overline{x}=\left(
x_{2},x_{3},...,x_{n}\right) .$ Suppose that the part $\Gamma \in C^{3}$ of
the boundary $\partial \Omega $ is: 
\begin{equation}
\left. 
\begin{array}{c}
\Gamma =\left\{ x:x_{1}=\omega \left( \overline{x}\right) \right\} \subset
\partial \Omega ,\text{ where }\overline{x}\in \left\{ \left\vert \overline{x%
}\right\vert <A_{1}\right\} , \\ 
\omega \in C^{3}\left( \left\vert \overline{x}\right\vert \leq A_{1}\right) 
\text{ and let }\Omega \subset \left\{ x_{1}>\omega \left( \overline{x}%
\right) ,\text{ }\overline{x}\in \left\{ \left\vert \overline{x}\right\vert
<A_{1}\right\} \right\} .%
\end{array}%
\right.  \label{5.1}
\end{equation}%
Change variables as 
\begin{equation}
\left( x_{1},\overline{x}\right) \Leftrightarrow \left( x_{1}^{\prime
}=x_{1}-\omega \left( \overline{x}/A_{1}\right) ,\overline{x}^{\prime }=%
\overline{x}/A_{1}\right) .  \label{5.2}
\end{equation}%
For brevity, we keep the same notation for $\left( x_{1}^{\prime },\overline{%
x}^{\prime }\right) $ and $\Omega $. Then by (\ref{5.1}) $\Gamma $ becomes: $%
\Gamma =\left\{ x:x_{1}=0,\left\vert \overline{x}\right\vert <1\right\} .$
Since the domain $\Omega $ is bounded, then it follows from (\ref{5.1}) and (%
\ref{5.2}) that we can assume that $\Omega \subset \left\{ x_{1}\in \left(
0,A_{2}\right) ,\left\vert \overline{x}\right\vert <1\right\} .$ Changing
variables again $x_{1}\Leftrightarrow x_{1}^{\prime }=x_{1}/\left(
2A_{2}\right) $ and again keeping the same notation for $x_{1}$ for brevity,
we assume below that%
\begin{equation}
\left. 
\begin{array}{c}
\Omega \subseteq \left\{ x:x_{1}\in \left( 0,1/2\right) ,\text{ }\left\vert 
\overline{x}\right\vert <1\right\} , \\ 
\Gamma =\left\{ x:x_{1}=0,\left\vert \overline{x}\right\vert <1\right\}
\subset \partial \Omega .%
\end{array}%
\right.  \label{5.3}
\end{equation}

Define the function $\psi \left( x\right) $ as%
\begin{equation}
\psi \left( x\right) =x_{1}+\frac{\left\vert \overline{x}\right\vert ^{2}}{2}%
+\frac{1}{4}.  \label{5.5}
\end{equation}%
Let $\lambda \geq 1$ and $\nu \geq 1$ be two large parameters. Define the
Carleman Weight Function $\varphi _{\lambda ,\nu }\left( x\right) $ as \cite[%
section 1 of chapter 4]{LRS}, \cite[formula (2.66)]{KL} 
\begin{equation}
\varphi _{\lambda ,\nu }\left( x\right) =\exp \left( 2\lambda \psi ^{-\nu
}\left( x\right) \right) .  \label{5.6}
\end{equation}%
Let $\gamma \in \left[ 0,1/2\right) $ be an arbitrary number from this
interval. Define the domains $G$, $G_{\alpha }$ as:%
\begin{equation}
G=\left\{ x:x_{1}>0,x_{1}+\frac{\left\vert \overline{x}\right\vert ^{2}}{2}+%
\frac{1}{4}<\frac{3}{4}\right\} =\left\{ x_{1}>0,\psi \left( x\right) <\frac{%
3}{4}\right\} ,  \label{5.7}
\end{equation}%
\begin{equation}
G_{\gamma }=\left\{ x:x_{1}>0,x_{1}+\frac{\left\vert \overline{x}\right\vert
^{2}}{2}+\frac{1}{4}<\frac{3}{4}-\gamma \right\} =\left\{ x_{1}>0,\psi
\left( x\right) <\frac{3}{4}-\gamma \right\} .  \label{5.8}
\end{equation}%
By (\ref{5.3}), (\ref{5.7}) and (\ref{5.8}) 
\begin{equation}
\left. 
\begin{array}{c}
G_{\gamma }\subset G\text{ for }\gamma \in \left( 0,1/2\right) \text{, }%
G_{0}=G, \\ 
G\subseteq \Omega \text{.}%
\end{array}%
\right.  \label{5.80}
\end{equation}%
It follows from (\ref{5.7}) and (\ref{5.8}) that the boundaries $\partial G$
and $\partial G_{\alpha }$ of the domains $G$ and $G_{\alpha }$ are%
\begin{equation}
\left. 
\begin{array}{c}
\partial G=\partial _{1}G\cup \partial _{2}G, \\ 
\partial _{1}G=\left\{ x_{1}=0,\left\vert \overline{x}\right\vert <1\right\}
=\Gamma , \\ 
\partial _{2}G=\left\{ x_{1}>0,\psi \left( x\right) =3/4\right\} , \\ 
\text{ }\partial G_{\gamma }=\partial _{1}G_{\gamma }\cup \partial
_{2}G_{\gamma }, \\ 
\partial _{1}G_{\gamma }=\left\{ x_{1}=0,\left\vert \overline{x}\right\vert
^{2}/2<1/2-\gamma \right\} , \\ 
\partial _{2}G_{\gamma }=\left\{ x_{1}>0,\psi \left( x\right) =3/4-\gamma
\right\} .%
\end{array}%
\right.  \label{5.9}
\end{equation}%
Hence, $\partial _{2}G_{\gamma }$ is such a part of the level surface of the
function $\psi \left( x\right) ,$ which is located in the half space $%
\left\{ x_{1}>0\right\} .$

\textbf{Remark 5.1.} \emph{Thus, we assume below that (\ref{5.3}) holds and
keep notations (\ref{5.5})-(\ref{5.9}). It follows from (\ref{5.1}) and the
transformation (\ref{5.2}) that the results below are applicable to quite
general domains }$\Omega $\emph{.}

\subsection{Carleman estimates}

\label{sec:5.2}

There exist two methods of proofs of Carleman estimates. The first one uses
symbols of partial differential operators, see, e.g. \cite[sections 8.3 and
8.4]{Horm}, \cite[Theorem 3.2.1]{Isakov}. Although this method is both
elegant and short, it assumes zero boundary conditions of involved
functions. Hence, it does not allow to incorporate non zero boundary
conditions. The second method uses space consuming derivations of pointwise
Carleman estimates. As soon as a pointwise estimate is derived, it is
integrated over an appropriate domain, Gauss formula results in boundary
integrals, which, in turn allow to use non-zero boundary conditions, see,
e.g. \cite{Ksurvey}, \cite[sections 2.3, 2.4]{KL}, \cite[section 1 of
chapter 4]{LRS}, \cite{Yam}. Hence, we formulate pointwise Carleman
estimates first and integrate them next.

Carleman estimates for elliptic operators are usually derived from their
counterparts for parabolic operators under the assumption of the
independence on $t$ of all involved functions. Keeping this in mind, there
are two types of Carleman estimates for elliptic operators. In the first
type only the tested function and its gradient are estimated. This kind of
estimates allow one to work with non-zero Dirichlet and Neumann boundary
conditions. In the second type, estimates of second order $x-$derivatives
are incorporated as well, although with a small parameter $1/\lambda .$ In
this case one can work only with the zero boundary conditions. These two
types of Carleman estimates are formulated below. It is well known that
Carleman estimates are formulated only for principal parts of Partial
Differential Operators and do not depend on their low order terms, see, e.g. 
\cite[Lemma 2.1.1]{KL}.

\textbf{Theorem 5.1} (the Carleman estimate of the first type) \cite[Lemma 3
in section 1 of chapter 4]{LRS}, \cite[Theorem 2.4.1]{KL}. \emph{Let the
domain }$\Omega $\emph{\ be as in (\ref{5.3}). Let }$L_{0}$\emph{\ be the
principal part of the elliptic operator in (\ref{3.6}). Assume that
conditions (\ref{3.5}) are satisfied. Then there exist sufficiently large
numbers }$\lambda _{0}=\lambda _{0}\left( \Omega ,a^{0},\mu \right) \geq 1$%
\emph{\ and }$\nu _{0}=\nu _{0}\left( G,a^{0},\mu \right) \geq 1$\emph{\ and
a number }$C=C\left( \Omega ,a^{0},\mu \right) >0,$ \emph{all three numbers
depending only on listed parameters, such that the following pointwise
Carleman estimate is valid with the CWF }$\varphi _{\lambda ,\nu }\left(
x\right) $\emph{\ defined in (\ref{5.6}):} \ 
\begin{equation}
\left. 
\begin{array}{c}
\left( L_{0}u\right) ^{2}\varphi _{\lambda ,\nu }\geq C\lambda \nu
\left\vert \nabla u\right\vert ^{2}\varphi _{\lambda ,\nu }+C\lambda ^{3}\nu
^{4}\psi ^{-2\nu -2}u^{2}\varphi _{\lambda ,\nu }+\func{div}U_{1}, \\ 
\left\vert U_{1}\right\vert \leq C\lambda ^{3}\nu ^{4}\psi ^{-2\nu -2}\left(
\left\vert \nabla u\right\vert ^{2}+u^{2}\right) \varphi _{\lambda ,\nu },
\\ 
\forall \lambda \geq \lambda _{0},\text{ }\forall \nu \geq \nu _{0},\text{ }%
\forall x\in \overline{G},\text{ }\forall u\in C^{2}\left( \overline{\Omega }%
\right) .%
\end{array}%
\right.  \label{5.10}
\end{equation}

In Theorem 5.2 we incorporate second order derivatives of the function $u$
in estimate (\ref{5.10}). It is convenient to set in this theorem $\nu =\nu
_{0}.$

\textbf{Theorem 5.2} (the Carleman estimate of the second type). \emph{%
Assume that conditions of Theorem 5.1 hold. Fix }$\nu =\nu _{0}\left( \Omega
,a^{0},\mu \right) .$\emph{\ Then the Carleman estimate (\ref{5.10}) can be
modified as:}%
\begin{equation}
\left. 
\begin{array}{c}
\left( L_{0}u\right) ^{2}\varphi _{\lambda ,\nu _{0}}\geq \left( C/\lambda
\right) \dsum\limits_{i,j=1}^{n}u_{x_{i}x_{j}}^{2}\varphi _{\lambda ,\nu
_{0}}+ \\ 
+C\lambda \left\vert \nabla u\right\vert ^{2}\varphi _{\lambda ,\nu
_{0}}+C\lambda ^{3}u^{2}\varphi _{\lambda ,\nu _{0}}+\func{div}U_{1}+\func{%
div}U_{2}, \\ 
\left\vert U_{1}\right\vert \leq C\lambda ^{3}\left( \left\vert \nabla
u\right\vert ^{2}+u^{2}\right) \varphi _{\lambda ,\nu _{0}},\text{ } \\ 
\left\vert U_{2}\right\vert \leq \left( C/\lambda \right) \left\vert \nabla
u\right\vert \dsum\limits_{i,j=1}^{n}\left\vert u_{x_{i}x_{j}}\right\vert
\varphi _{\lambda ,\nu _{0}}, \\ 
\forall \lambda \geq \lambda _{0},\text{ }\forall x\in \overline{\Omega },%
\text{ }\forall u\in C^{3}\left( \overline{\Omega }\right) .%
\end{array}%
\right.  \label{5.11}
\end{equation}

\textbf{Proof.} By Theorem 5.1 we need to prove only the involvement of the
second derivatives and the estimate in the fourth line of (\ref{5.11}). We
have:%
\begin{equation}
\left( L_{0}u\right) ^{2}\varphi _{\lambda ,\nu
_{0}}^{2}=\dsum\limits_{i,j=1}^{n}\dsum%
\limits_{k,s=1}^{n}a_{ij}a_{ks}u_{x_{i}x_{j}}u_{x_{k}x_{s}}\varphi _{\lambda
,\nu _{0}}^{2}.  \label{5.12}
\end{equation}%
In addition, we have:%
\begin{equation}
\left. 
\begin{array}{c}
a_{ij}a_{ks}u_{x_{i}x_{j}}u_{x_{k}x_{s}}\varphi _{\lambda ,\nu _{0}}^{2}= \\ 
=\left( a_{ij}a_{ks}u_{x_{i}x_{j}}u_{x_{k}}\varphi _{\lambda ,\nu
_{0}}^{2}\right) _{x_{s}}-a_{ij}a_{ks}u_{x_{i}x_{j}x_{s}}u_{x_{k}}\varphi
_{\lambda ,\nu _{0}}^{2}-\left( a_{ij}a_{ks}\varphi _{\lambda ,\nu
_{0}}^{2}\right) _{x_{s}}u_{x_{i}x_{j}}u_{x_{k}}= \\ 
=\left( a_{ij}a_{ks}u_{x_{i}x_{j}}u_{x_{k}}\varphi _{\lambda ,\nu
_{0}}^{2}\right) _{x_{s}}-\left( a_{ij}a_{ks}\varphi _{\lambda ,\nu
_{0}}^{2}\right) _{x_{s}}u_{x_{i}x_{j}}u_{x_{k}}+ \\ 
+\left( -a_{ij}a_{ks}u_{x_{i}x_{s}}u_{x_{k}}\varphi _{\lambda ,\nu
_{0}}^{2}\right) _{x_{j}}+\left( a_{ij}a_{ks}u_{x_{i}x_{s}}u_{x_{k}}\varphi
_{\lambda ,\nu _{0}}^{2}\right)
_{x_{j}}+a_{ij}a_{ks}u_{x_{i}x_{s}}u_{x_{k}x_{j}}\varphi _{\lambda ,\nu
_{0}}^{2}.%
\end{array}%
\right.  \label{5.13}
\end{equation}%
It was proven in \cite[formula (6.12) of Chapter 2]{Lad} that 
\begin{equation}
\dsum\limits_{i,j=1}^{n}\dsum%
\limits_{k,s=1}^{n}a_{ij}a_{ks}u_{x_{i}x_{s}}u_{x_{k}x_{j}}\geq \mu
^{2}\dsum\limits_{i,j=1}^{n}u_{x_{i}x_{j}}^{2}.  \label{5.14}
\end{equation}%
Hence, (\ref{5.5}), (\ref{5.6}), (\ref{5.12})-(\ref{5.14}) and
Cauchy-Schwarz inequality imply:%
\begin{equation}
\left. 
\begin{array}{c}
\left( L_{0}u\right) ^{2}\varphi _{\lambda ,\nu _{0}}\geq
C\dsum\limits_{i,j=1}^{n}u_{x_{i}x_{j}}^{2}\varphi _{\lambda ,\nu
_{0}}-C\lambda ^{2}\left\vert \nabla u\right\vert ^{2}\varphi _{\lambda ,\nu
_{0}}+\func{div}\widetilde{U}_{2}, \\ 
\left\vert \widetilde{U}_{2}\right\vert \leq C\left\vert \nabla u\right\vert
\dsum\limits_{i,j=1}^{n}\left\vert u_{x_{i}x_{j}}\right\vert \varphi
_{\lambda ,\nu _{0}}.%
\end{array}%
\right.  \label{5.15}
\end{equation}%
Divide both formulas (\ref{5.15}) by $d\cdot \lambda $ with an appropriate
number $d=d\left( \Omega ,a^{0},\mu \right) >0$ and sum up with (\ref{5.10})
setting there $\nu =\nu _{0}.$ Then we obtain (\ref{5.11}). $\ \square $

\textbf{Remark 5.2}. \emph{It is clear from (\ref{5.10}) and density
arguments that the integration of these pointwise Carleman estimates over
the domain }$G$\emph{\ makes the resulting integral estimate valid for all
functions }$u\in H^{2}\left( G\right) $\emph{, and in the case of (\ref{5.11}%
) for all functions }$u\in \left\{ u\in H^{2}\left( G\right) :\nabla u=0%
\text{ on }\partial G\right\} $\emph{.}

\section{H\"{o}lder and Lipschitz Stability Estimates for the Unique
Continuation Problem (\protect\ref{3.0})-(\protect\ref{3.4})}

\label{sec:6}

Theorems of this section address, within the TFD framework, the first long
standing open question formulated in Introduction.

\subsection{The first H\"{o}lder stability estimate}

\label{sec:6.1}

\textbf{Theorem 6.1 }(the first H\"{o}lder stability estimate)\textbf{.} 
\emph{Let }$M_{1},M_{2}>0$\emph{\ be two numbers. Consider the ball }%
\begin{equation}
D\left( M_{1}\right) =\left\{ y\in \mathbb{R}^{N}:\left\vert
y_{i}\right\vert <M_{1},i=1,...,N\right\} ,  \label{6.0}
\end{equation}%
\emph{where a special role of the vector }$y\in \mathbb{R}^{N}$\emph{\ is
explained in the paragraph below (\ref{3.1}).\ Let}%
\begin{equation}
\left\Vert F\left( y,x,t\right) \right\Vert _{C^{2}\left( \overline{D\left(
M_{1}\right) }\right) \times C\left( \overline{\Omega }_{h,T}\right) }\leq
M_{2}.  \label{6.1}
\end{equation}%
\emph{Let condition (\ref{1.1}) be valid. Let the domain }$G$\emph{\ be the
one defined in (\ref{5.7}) and the domain }$\Omega $ \emph{be the one in (%
\ref{5.3}). Assume that conditions (\ref{3.2}) and (\ref{3.3}) hold, where }$%
\Omega $\emph{\ is replaced with }$G.$ \emph{Given notation (\ref{3.0}), let
two vector functions }%
\begin{equation}
u_{1}\left( x,t\right) ,u_{2}\left( x,t\right) \in C^{2}\left( \overline{G}%
_{h,T}\right)  \label{6.002}
\end{equation}%
\emph{\ are solutions of equation (\ref{3.1}) in }$G_{h,T}$\emph{\ with the
lateral Cauchy data (\ref{3.4}) at }$\Gamma _{h,T}.$ \emph{Assume that }%
\begin{equation}
\left\Vert u_{i}\right\Vert _{C^{2}\left( \overline{G}_{h,T}\right) }\leq
M_{1},\text{ }i=1,2.  \label{6.2}
\end{equation}%
\emph{\ Denote }%
\begin{equation}
\overline{u}=u_{1}-u_{2}\emph{,\ }\overline{g}_{0}=g_{1,0}-g_{2,0},\emph{\ }%
\overline{g}_{1}=g_{1,1}-g_{2,1}.  \label{6.03}
\end{equation}%
\emph{\ Let }$\delta \in \left( 0,1\right) $ \emph{be a number. Assume that }%
\begin{equation}
\left\Vert \overline{g}_{0}\right\Vert _{H^{1}\left( \Gamma _{h,T}\right)
}<\delta \text{ and }\left\Vert \overline{g}_{1}\right\Vert _{L_{2}\left(
\Gamma _{h,T}\right) }<\delta .  \label{6.3}
\end{equation}%
\emph{Norms in (\ref{6.3}) are understood as in (\ref{2.30}). Choose an
arbitrary number }$\varepsilon \in \left( 0,1/6\right) .$\emph{\ Then there
exists a sufficiently small number }$\delta _{1}=\delta _{1}\left(
M_{1},M_{2},\mu ,G,T,h_{0},\varepsilon \right) \in \left( 0,1\right) $\emph{%
\ such that the following H\"{o}lder stability estimate holds:}%
\begin{equation}
\left\Vert \overline{u}\right\Vert _{H^{1}\left( G_{3\varepsilon
,h,T}\right) }\leq C_{1}\left( 1+\left\Vert \overline{u}\right\Vert
_{H^{1}\left( G_{h,T}\right) }\right) \delta ^{\rho _{1}},\text{ }\forall
\delta \in \left( 0,\delta _{1}\right) ,  \label{6.4}
\end{equation}%
\emph{where the number }$\rho _{1}=\rho _{1}\left( M_{1},M_{2},\mu
,G,T,h_{0},\varepsilon \right) \in \left( 0,1\right) $\emph{\ and the number 
}

$C_{1}=C_{1}\left( M_{1},M_{2},\mu ,G,T,h_{0},\varepsilon \right) >0.$\emph{%
\ Numbers }$\delta _{1},\rho _{1}$\emph{\ and }$C_{1}$\emph{\ depend only on
listed parameters. In addition, problem (\ref{3.0}), (\ref{3.1}), (\ref{3.4}%
) has at most one solution }$u\in C^{3}\left( \overline{G}_{h,T}\right) .$

\textbf{Remark 6.1.} \emph{Below }$C_{1}>0$\emph{\ denotes different numbers
depending only on the above listed parameters.}

\textbf{Proof of Theorem 6.1}. It follows from (\ref{3.4}), (\ref{6.002})
and (\ref{6.03}) that norms in (\ref{6.3}) make sense. Using (\ref{2.7}), (%
\ref{2.8}) and (\ref{6.2}), we obtain%
\begin{equation}
\left. 
\begin{array}{c}
\max_{t_{s}\in Y}\left\Vert \left( \dint\limits_{0}^{t_{s}}\left\vert \nabla
u_{i}\left( x,\tau \right) \right\vert d\tau \right) _{h}\right\Vert
_{C\left( \overline{\Omega }\right) }\leq C_{1}M_{1}, \\ 
\max_{t_{s}\in Y}\left\Vert \left( \dint\limits_{0}^{t_{s}}\left\vert
u_{i}\left( x,\tau \right) \right\vert d\tau \right) _{h}\right\Vert
_{C\left( \overline{\Omega }\right) }\leq C_{1}M_{1} \\ 
\max_{t_{s}\in Y}\left\Vert \left( \dint\limits_{0}^{t_{s}}\left\vert
\partial _{h,t}u_{i}\left( x,\tau \right) \right\vert d\tau \right)
_{h}\right\Vert _{C\left( \overline{\Omega }\right) }\leq C_{1}M_{1},\text{ }
\\ 
i=1,2,%
\end{array}%
\right.  \label{6.02}
\end{equation}%
where integrals are understood in terms of (\ref{2.7}) and (\ref{2.8}). Note
that by (\ref{5.5}), (\ref{5.7}) and (\ref{5.8})%
\begin{equation}
G_{3\varepsilon }\neq \varnothing \text{ and }\psi \left( x\right) \in
\left( 1/4,3/4-3\varepsilon \right) \text{ in }G_{3\varepsilon }.
\label{6.5}
\end{equation}%
Subtract equation (\ref{3.1}) for the vector $u_{2}$ from the same equation
for $u_{1}.$ Using (\ref{3.0}) and the finite increment formula, we obtain
for the vector function $\overline{u}:$%
\begin{equation}
\left. 
\begin{array}{c}
\partial _{h,t}\overline{u}\left( x,t_{s}\right)
=\dsum\limits_{i,j=1}^{n}b_{i,j}\left( x,t_{s}\right) \overline{u}%
_{x_{i}x_{j}}\left( x,t_{s}\right) + \\ 
+\dsum\limits_{j=1}^{n}d_{j}^{\left( 1\right) }\left( x,t_{s}\right) 
\overline{u}_{x_{j}}\left( x,t_{s}\right) +d_{0}^{\left( 1\right) }\left(
x,t_{s}\right) \overline{u}\left( x,t_{s}\right) + \\ 
+\dsum\limits_{j=1}^{n}d_{j}^{\left( 2\right) }\left( x,t_{s}\right) \left(
\dint\limits_{0}^{t_{s}}\overline{u}_{x_{j}}\left( x,\tau \right) d\tau
\right) _{h}+d_{0}^{\left( 2\right) }\left( x,t_{s}\right) \left(
\dint\limits_{0}^{t_{s}}\overline{u}\left( x,\tau \right) d\tau \right) _{h}+
\\ 
+d_{0}^{\left( 3\right) }\left( x,t_{s}\right) \left(
\dint\limits_{0}^{t_{s}}\partial _{h,t}\overline{u}\left( x,\tau \right)
d\tau \right) _{h},\text{ }x\in G,t_{s}\in Y, \\ 
\overline{u}\mid _{\Gamma _{h,T}}=\overline{g}_{0}\left( x,t_{s}\right) ,%
\text{ }\overline{u}_{x_{1}}\mid _{\Gamma _{h,T}}=-\overline{g}_{1}\left(
x,t_{s}\right) ,\text{ }t_{s}\in Y,,%
\end{array}%
\right.  \label{6.6}
\end{equation}%
where $\Gamma $ is given in (\ref{5.3}). It follows from (\ref{6.0})-(\ref%
{6.2}) that in (\ref{6.6}) vector functions%
\begin{equation}
\left. 
\begin{array}{c}
b_{i,j},d_{r}^{\left( s\right) }\in C^{1}\left( \overline{\Omega }%
_{h,T}\right) ;\text{ }\left\Vert b_{i,j}\right\Vert _{C^{1}\left( \overline{%
\Omega }_{h,T}\right) },\left\Vert d_{r}^{\left( s\right) }\right\Vert
_{C^{1}\left( \overline{\Omega }_{h,T}\right) }\leq M_{1}, \\ 
\text{ }i,j=1,...,n;\text{ }r=0,...,n;\text{ }s=1,2,3.%
\end{array}%
\right.  \label{6.7}
\end{equation}%
In addition, it follows from (\ref{3.2}) and (\ref{3.3}) that 
\begin{equation}
\left. 
\begin{array}{c}
b_{i,j}\left( x,t_{s}\right) =b_{j,i}\left( x,t_{s}\right) ,\text{ }\left(
x,t_{s}\right) \in G_{h,T}, \\ 
\mu \left\vert \xi \right\vert ^{2}\leq
\dsum\limits_{i,j=1}^{n}b_{i,j}\left( x,t_{s}\right) \xi _{i}\xi _{j},\text{ 
}\left( x,t_{s}\right) \in G_{h,T}.%
\end{array}%
\right.  \label{6.8}
\end{equation}

Denote%
\begin{equation}
L^{\left( s\right) }\left( \overline{u}\left( x,t_{s}\right) \right)
=\dsum\limits_{i,j=1}^{n}b_{i,j}\left( x,t_{s}\right) \overline{u}%
_{x_{i}x_{j}}\left( x,t_{s}\right) ,\text{ }s=0,...,k,x\in \Omega .
\label{6.9}
\end{equation}%
It follows from (\ref{2.4})-(\ref{2.6}) that the function $\overline{u}%
\left( x,t_{s}\right) /\left( 2h\right) $ is involved two times in the left
hand side of (\ref{6.6}) for each $s=0,...,k$, also, see Remark 2.1. Hence,
using (\ref{2.4})-(\ref{2.8}), (\ref{6.02}), (\ref{6.6}), (\ref{6.7}) and (%
\ref{6.9}), we obtain the following inequalities:%
\begin{equation}
\left\vert L^{\left( s\right) }\left( \overline{u}\left( x,t_{s}\right)
\right) \right\vert \leq C_{1}\dsum\limits_{j=0}^{k}\left( \left\vert \nabla 
\overline{u}\left( x,t_{j}\right) \right\vert +\left\vert \overline{u}\left(
x,t_{j}\right) \right\vert \right) ,\text{ }x\in G;\text{ }t_{j},t_{s}\in Y.
\label{6.10}
\end{equation}%
Consider a cut-off function $\chi _{\varepsilon }\left( x\right) $ such that 
\begin{equation}
\chi _{\varepsilon }\left( x\right) \in C^{2}\left( \overline{G}\right) ,%
\text{ }\chi _{\varepsilon }\left( x\right) =\left\{ 
\begin{array}{c}
1\text{ in }G_{2\varepsilon }, \\ 
0\text{ in }G\diagdown G_{\varepsilon }, \\ 
\text{ between }0\text{ and }1\text{ in }G_{\varepsilon }\diagdown
G_{2\varepsilon }.%
\end{array}%
\right. \text{ }  \label{6.11}
\end{equation}%
The existence of such functions is well known from the Analysis course.
Denote 
\begin{equation}
v\left( x,t_{i}\right) =\chi _{\varepsilon }\left( x\right) \overline{u}%
\left( x,t_{i}\right) ,\text{ }\left( x,t_{i}\right) \in G_{h,T}.
\label{6.12}
\end{equation}%
It follows from (\ref{5.9}) and (\ref{6.11}) and (\ref{6.12}) that 
\begin{equation}
v\left( x,t_{i}\right) =0\text{ in a small neighborhood of }\partial _{2}G.
\label{6.13}
\end{equation}%
Multiply both sides of (\ref{6.10}) by $\chi _{\varepsilon }\left( x\right)
, $ keeping in mind that by (\ref{6.11}) $\chi _{\varepsilon }\left(
x\right) \geq 0$ in $G$. Use:%
\begin{equation*}
\left. 
\begin{array}{c}
\chi _{\varepsilon }\overline{u}_{x_{i}}=v_{x_{i}x_{i}}-\left( \chi
_{\varepsilon }\right) _{x_{i}}\overline{u}=v_{x_{i}}-\left( \chi
_{\varepsilon }\right) _{x_{i}}\overline{u}, \\ 
\chi _{\varepsilon }\overline{u}_{x_{i}x_{j}}=v_{x_{i}x_{j}}-\left( \chi
_{\varepsilon }\right) _{x_{j}}\overline{u}_{x_{i}}-\left( \chi
_{\varepsilon }\right) _{x_{i}}\overline{u}_{x_{j}}-\left( \chi
_{\varepsilon }\right) _{x_{i}x_{j}}\overline{u}.%
\end{array}%
\right.
\end{equation*}%
Hence, using the last line of (\ref{6.6}) and (\ref{6.12}), we obtain%
\begin{equation}
\left. 
\begin{array}{c}
\left\vert L^{\left( s\right) }\left( v\left( x,t_{s}\right) \right)
\right\vert \leq C_{1}\dsum\limits_{j=0}^{k}\left( \left\vert \nabla v\left(
x,t_{j}\right) \right\vert +\left\vert v\left( x,t_{j}\right) \right\vert
\right) \\ 
+C_{1}\left( 1-\chi _{\varepsilon }\left( x\right) \right) \left(
\dsum\limits_{j=0}^{s}\left\vert \nabla \overline{u}\left( x,t_{j}\right)
\right\vert +\dsum\limits_{j=0}^{k}\left\vert \overline{u}\left(
x,t_{j}\right) \right\vert \right) ,\text{ }x\in G;\text{ }t_{j},t_{s}\in Y,
\\ 
v\mid _{\Gamma _{h,T}}=\chi _{\varepsilon }\left( x\right) \overline{g}%
_{0}\left( x,t_{s}\right) ,\text{ }v_{x_{1}}\mid _{\Gamma _{h,T}}=-\left(
\left( \chi _{\varepsilon }\right) _{x_{1}}\overline{g}_{0}+\chi
_{\varepsilon }\overline{g}_{1}\right) \left( x,t_{s}\right) .\text{ }%
\end{array}%
\right.  \label{6.14}
\end{equation}

Square both sides of each inequality (\ref{6.14}), use Cauchy-Schwarz
inequality and then multiply both sides of the resulting inequality by the
function $\varphi _{\lambda ,\nu _{0}}\left( x\right) ,$ where $\varphi
_{\lambda ,\nu _{0}}\left( x\right) $ is the CWF (\ref{5.6}) at $\nu =\nu
_{0},$ where $\nu _{0}$ is the number chosen in Theorem 5.1. We obtain%
\begin{equation}
\left. 
\begin{array}{c}
\left[ L^{\left( s\right) }\left( v\left( x,t_{s}\right) \right) \right]
^{2}\varphi _{\lambda ,\nu _{0}}\left( x\right) \leq \\ 
\leq C_{1}\dsum\limits_{i=0}^{k}\left( \left\vert \nabla v\left(
x,t_{j}\right) \right\vert ^{2}+\left\vert v\left( x,t_{j}\right)
\right\vert ^{2}\right) \varphi _{\lambda ,\nu _{0}}\left( x\right) + \\ 
+C_{1}\left( 1-\chi _{\varepsilon }\left( x\right) \right)
\dsum\limits_{i=0}^{k}\left( \left\vert \nabla u\left( x,t_{j}\right)
\right\vert ^{2}+\left\vert u\left( x,t_{j}\right) \right\vert ^{2}\right)
\varphi _{\lambda ,\nu _{0}}\left( x\right) ,\text{ } \\ 
x\in G;\text{ }t_{j},t_{s}\in Y.%
\end{array}%
\right.  \label{6.15}
\end{equation}%
It follows from (\ref{6.7})-(\ref{6.9}) that we can apply the pointwise
Carleman estimate (\ref{5.10}) to the left hand side of (\ref{6.15}).
Integrate each of those estimates over the domain $G$, using Gauss formula,
the second line of (\ref{5.10}) and (\ref{6.13}). Then sum up the resulting
estimates with respect to the index $s=0,...,k.$ We also note that by (\ref%
{5.3})-(\ref{5.7}) 
\begin{equation}
\max_{\overline{G}}\varphi _{\lambda ,\nu _{0}}\left( x\right) =\exp \left(
2\lambda \cdot 4^{\nu _{0}}\right) .  \label{6.140}
\end{equation}%
Let $\lambda _{0}$ be the number of Theorem 5.1. We obtain for all $\lambda
\geq \lambda _{0}$%
\begin{equation}
\left. 
\begin{array}{c}
\exp \left( 2\lambda \cdot 4^{\nu _{0}}\right) \left( \left\Vert \overline{g}%
_{0}\right\Vert _{H^{1}\left( \Gamma _{h,T}\right) }^{2}+\left\Vert 
\overline{g}_{1}\right\Vert _{L_{2}\left( \Gamma _{h,T}\right) }^{2}\right) +
\\ 
+\dint\limits_{G}\left( 1-\chi _{\varepsilon }\left( x\right) \right) \left(
\dsum\limits_{j=0}^{k}\left( \left\vert \nabla \overline{u}\left(
x,t_{j}\right) \right\vert ^{2}+\left\vert \overline{u}\left( x,t_{j}\right)
\right\vert ^{2}\right) \right) \varphi _{\lambda ,\nu _{0}}\left( x\right)
dx+ \\ 
+\dint\limits_{G}\dsum\limits_{j=0}^{k}\left( \left\vert \nabla v\left(
x,t_{j}\right) \right\vert ^{2}+\left\vert v\left( x,t_{j}\right)
\right\vert ^{2}\right) \varphi _{\lambda ,\nu _{0}}\left( x\right) dx\geq
\\ 
\geq C_{1}\lambda \dint\limits_{G}\dsum\limits_{j=0}^{k}\left( \left\vert
\nabla v\left( x,t_{j}\right) \right\vert ^{2}+\lambda ^{2}\left\vert
v\left( x,t_{j}\right) \right\vert ^{2}\right) \varphi _{\lambda ,\nu
_{0}}\left( x\right) dx.%
\end{array}%
\right.  \label{6.16}
\end{equation}%
Note that by (\ref{6.11}) $1-\chi _{\varepsilon }\left( x\right) =0$ in $%
G_{2\varepsilon }$ and by (\ref{6.3}) $\left\Vert \overline{g}%
_{0}\right\Vert _{H^{1}\left( \Gamma _{h,T}\right) }^{2}+\left\Vert 
\overline{g}_{1}\right\Vert _{L_{2}\left( \Gamma _{h,T}\right) }^{2}<2\delta
^{2}.$ Choose $\lambda _{1}=\lambda _{1}\left( M_{1},M_{2},\mu ,\Omega
,T,h_{0},\varepsilon \right) \geq \lambda _{0}$ depending only on listed
parameters such that $C_{1}\lambda _{1}/2>1.$ Hence, using (\ref{6.16}), we
obtain%
\begin{equation}
\left. 
\begin{array}{c}
\exp \left( 2\cdot 4^{\nu _{0}}\cdot \lambda \right) \delta ^{2}+ \\ 
+\dint\limits_{G\diagdown G_{2\varepsilon }}\dsum\limits_{j=0}^{k}\left(
\left\vert \nabla \overline{u}\left( x,t_{j}\right) \right\vert
^{2}+\left\vert \overline{u}\left( x,t_{j}\right) \right\vert ^{2}\right)
\varphi _{\lambda ,\nu _{0}}\left( x\right) dx\geq \\ 
\geq C_{1}\lambda \dint\limits_{G}\dsum\limits_{j=0}^{k}\left( \left\vert
\nabla v\left( x,t_{j}\right) \right\vert ^{2}+\lambda ^{2}\left\vert
v\left( x,t_{j}\right) \right\vert ^{2}\right) \varphi _{\lambda ,\nu
_{0}}\left( x\right) dx\geq \\ 
\geq C_{1}\lambda \dint\limits_{G_{3\varepsilon
}}\dsum\limits_{j=0}^{k}\left( \left\vert \nabla \overline{u}\left(
x,t_{j}\right) \right\vert ^{2}+\lambda ^{2}\left\vert \overline{u}\left(
x,t_{j}\right) \right\vert ^{2}\right) \varphi _{\lambda ,\nu _{0}}\left(
x\right) dx,\text{ }\forall \lambda \geq \lambda _{1}.%
\end{array}%
\right.  \label{6.17}
\end{equation}%
The last line of (\ref{6.17}) is less or equal than its third line since the
first line of (\ref{5.80}) implies that $G_{3\varepsilon }\subset G.$ Also,
we use $\overline{u}$ instead of $v$ in the last line of (\ref{6.17}) due to
Next, by (\ref{5.7}) and (\ref{5.8})%
\begin{equation*}
\left. 
\begin{array}{c}
1/4<\psi \left( x\right) <3/4-3\varepsilon \text{ in }G_{3\varepsilon }, \\ 
3/4-2\varepsilon \text{ }<\psi \left( x\right) <3/4,\text{ in }G\diagdown
G_{2\varepsilon }.%
\end{array}%
\right.
\end{equation*}%
Hence, using (\ref{5.6}), we obtain 
\begin{equation}
\left. 
\begin{array}{c}
\varphi _{\lambda ,\nu _{0}}\left( x\right) \geq \exp \left[ 2\lambda \left(
3/4-3\varepsilon \right) ^{-\nu _{0}}\right] \text{ in }G_{3\varepsilon },
\\ 
\varphi _{\lambda ,\nu _{0}}\left( x\right) \leq \exp \left[ 2\lambda \left(
3/4-2\varepsilon \right) ^{-\nu _{0}}\right] \text{ in }G\diagdown
G_{2\varepsilon }.%
\end{array}%
\right.  \label{6.18}
\end{equation}%
Denote%
\begin{equation*}
\xi =\xi \left( M_{1},M_{2},\mu ,\Omega ,T,h_{0},\varepsilon \right) =\left[
\left( 3/4-3\varepsilon \right) ^{-\nu _{0}}-\left( 3/4-2\varepsilon \right)
^{-\nu _{0}}\right] >0.
\end{equation*}%
Hence, (\ref{6.17}) and (\ref{6.18}) lead to%
\begin{equation}
\left\Vert \overline{u}\right\Vert _{H^{1}\left( G_{3\varepsilon
,h,T}\right) }^{2}\leq C_{1}\exp \left( 2\lambda \cdot 4^{\nu _{0}}\right)
\delta ^{2}+C_{1}e^{-2\lambda \xi }\left\Vert \overline{u}\right\Vert
_{H^{1}\left( G_{h,T}\right) }^{2},\text{ }\forall \lambda \geq \lambda _{1}.
\label{6.19}
\end{equation}%
Choose $\lambda =\lambda \left( \delta \right) $ such that 
\begin{equation*}
\exp \left( 2\lambda \cdot 4^{\nu _{0}}\right) \delta ^{2}=e^{-2\lambda \xi
}.
\end{equation*}%
Hence, 
\begin{equation}
\lambda =\ln \left( \frac{1}{\delta ^{\theta }}\right) ,\text{ }\theta =%
\frac{1}{4^{\nu _{0}}+\xi }.  \label{6.20}
\end{equation}%
Hence, we should have $\delta \in \left( 0,\delta _{1}\right) ,$ where the
number $\delta _{1}=\delta _{1}\left( M_{1},M_{2},\mu ,\Omega
,T,h_{0},\varepsilon \right) \in \left( 0,1\right) $ is so small that $\ln
\left( \delta _{1}^{-\gamma }\right) \geq \lambda _{1}.$ Using (\ref{6.18})
and (\ref{6.18}), we obtain 
\begin{equation}
\left\Vert \overline{u}\right\Vert _{H^{1}\left( G_{3\varepsilon
,h,T}\right) }\leq C_{1}\left( 1+\left\Vert \overline{u}\right\Vert
_{H^{1}\left( G_{h,T}\right) }\right) \delta ^{\rho _{1}},\text{ }\forall
\delta \in \left( 0,\delta _{1}\right) ,  \label{6.21}
\end{equation}%
\begin{equation}
\rho _{1}=\frac{\xi }{\left( 4^{\nu _{0}}+\xi \right) }\in \left( 0,1\right)
.\text{ }  \label{6.22}
\end{equation}%
The target H\"{o}lder stability estimate (\ref{6.4}) follows immediately
from (\ref{6.21}) and (\ref{6.22}).

To prove uniqueness, we set $\delta =0.$ Since the number $\varepsilon \in
\left( 0,1/6\right) $ is an arbitrary one, then (\ref{6.4}) implies that $%
\overline{u}\left( x,t\right) =0$ in the entire domain $G_{h,T}.$ $\square $

\subsection{The second H\"{o}lder stability estimate}

\label{sec:6.2}

While Theorem 6.1 is about an estimate of the $H^{1}\left( G_{3\varepsilon
,h,T}\right) -$norm of the function $\overline{u}\left( x,t\right) ,$ we now
estimate in this subsection a stronger $H^{2}\left( G_{3\varepsilon
,h,T}\right) -$norm of this function. To do this, we use the Carleman
estimate of Theorem 5.2 instead of Theorem 5.1.

\textbf{Theorem 6.2 }(the second H\"{o}lder stability estimate)\textbf{. }%
\emph{Assume that all conditions of Theorem 6.1 hold, except that now (\ref%
{6.002}) is replaced with }%
\begin{equation}
u_{1}\left( x,t\right) ,u_{2}\left( x,t\right) \in C^{2}\left( \overline{G}%
_{h,T}\right)  \label{6.230}
\end{equation}%
\emph{and} \emph{(\ref{6.3}) is replaced with estimates in stronger norms\ }%
\begin{equation}
\left\Vert \overline{g}_{0}\right\Vert _{H^{2}\left( \Gamma _{h,T}\right)
}<\delta \text{ and }\left\Vert \overline{g}_{1}\right\Vert _{H^{2}\left(
\Gamma _{h,T}\right) }<\delta ,  \label{6.23}
\end{equation}%
\emph{see Remarks 2.1.} \emph{Norms in (\ref{6.23}) are understood as in (%
\ref{2.30}). Keep notations (\ref{6.03}). Choose an arbitrary number }$%
\varepsilon \in \left( 0,1/6\right) .$ \emph{Then there exists a
sufficiently small number }$\delta _{2}=\delta _{2}\left( M_{1},M_{2},\mu
,G,T,h_{0},\varepsilon \right) \in \left( 0,1\right) $\emph{\ and a number }$%
\rho _{2}=\rho _{2}\left( M_{1},M_{2},\mu ,G,T,h_{0},\varepsilon \right) \in
\left( 0,1\right) $ \emph{such that the following H\"{o}lder stability
estimate holds:}%
\begin{equation}
\left\Vert \overline{u}\right\Vert _{H^{2}\left( G_{3\varepsilon
,h,T}\right) }\leq C_{1}\left( 1+\left\Vert \overline{u}\right\Vert
_{H^{1}\left( G_{h,T}\right) }\right) \delta ^{\rho _{2}},\text{ }\forall
\delta \in \left( 0,\delta _{2}\right) ,  \label{6.24}
\end{equation}%
\emph{where the number }$C_{1}=C_{1}\left( M_{1},M_{2},\mu
,G,T,h_{0},\varepsilon \right) >0.$\emph{\ Numbers }$\delta _{2},\rho _{2}$%
\emph{\ and }$C_{1}$\emph{\ depend only on listed parameters. }

\textbf{Proof.} By (\ref{6.230}) norms in (\ref{6.23}) make sense. Keeping
in mind (\ref{5.3}) and (\ref{5.9}), consider the following function $%
\widetilde{u}\left( x,t\right) $ in $G_{h,T}:$%
\begin{equation}
\widetilde{u}\left( x,t_{s}\right) =\overline{u}\left( x,t_{s}\right) -%
\overline{g}_{0}\left( \overline{x},t_{s}\right) +x_{1}\overline{g}%
_{1}\left( \overline{x},t_{s}\right) ,\left( x,t_{s}\right) \in G_{h,T}.
\label{6.25}
\end{equation}%
Then the last line of (\ref{6.6}) implies:%
\begin{equation}
\widetilde{u}\mid _{\Gamma _{h,T}}=\widetilde{u}_{x_{1}}\mid _{\Gamma
_{h,T}}=0.  \label{6.26}
\end{equation}%
Taking from (\ref{6.25}) $\overline{u}=\widetilde{u}+\overline{g}_{0}+x_{1}%
\overline{g}_{1}$, substituting this in the first four lines of (\ref{6.6})
and then turning the resulting equation in the inequality, we obtain
similarly with (\ref{6.10})%
\begin{equation}
\left. 
\begin{array}{c}
\left\vert L^{\left( s\right) }\left( \widetilde{u}\left( x,t_{s}\right)
\right) \right\vert \leq C_{1}\dsum\limits_{j=0}^{k}\left( \left\vert \nabla 
\widetilde{u}\left( x,t_{j}\right) \right\vert +\left\vert \widetilde{u}%
\left( x,t_{j}\right) \right\vert \right) + \\ 
+C_{1}\dsum\limits_{\left\vert \alpha \right\vert \leq
2}\dsum\limits_{j=0}^{k}\dsum\limits_{i=0}^{1}\left\vert D_{x}^{\alpha }%
\overline{g}_{i}\left( x,t_{j}\right) \right\vert ,\text{ }x\in G;\text{ }%
t_{j},t_{s}\in Y.%
\end{array}%
\right.  \label{6.27}
\end{equation}%
Next, we proceed similarly with the proof of Theorem 6.1, except that we
apply the Carleman estimate \bigskip (\ref{5.11}) of Theorem 5.2 instead of (%
\ref{5.10}) of Theorem 5.1. It follows from the third and fourth lines of (%
\ref{5.11}), (\ref{6.13}) and (\ref{6.26}) that boundary terms will not
occur when using Gauss formula, also, see Remark 5.2. Then, using (\ref{6.23}%
), we obtain a direct analog of (\ref{6.21}), (\ref{6.22}), in which,
however, norms $\left\Vert \overline{u}\right\Vert _{H^{1}\left(
G_{h,T}\right) }$ and $\left\Vert \overline{u}\right\Vert _{H^{1}\left(
G_{3\varepsilon ,h,T}\right) }$ are replaced with $\left\Vert \widetilde{u}%
\right\Vert _{H^{2}\left( G_{h,T}\right) }$ and $\left\Vert \widetilde{u}%
\right\Vert _{H^{2}\left( G_{3\varepsilon ,h,T}\right) }$ respectively.
Next, the use of (\ref{6.25}) leads to (\ref{6.24}). $\square $

\subsection{Lipschitz stability estimates}

\label{sec:6.3}

We now replace the domain $G$ of Theorems 6.1 and 6.2 with the domain $%
\Omega $ and recall that by (\ref{5.80}) $G\subseteq \Omega $. We also
assume now that the lateral Cauchy data are given on the entire lateral
boundary $S_{h,T}$ of the semi-discrete time cylinder $\Omega _{h,T}$
instead of the part $\Gamma _{h,T}$ of $S_{h,T},$ see (\ref{2.2})-(\ref%
{2.210}) and (\ref{5.3}).

\textbf{Theorem 6.3} (the first Lipschitz stability estimate). \emph{Assume
that conditions (\ref{1.1}), (\ref{2.31}), (\ref{5.3}), (\ref{6.0}) and (\ref%
{6.1}) hold. Given notation (\ref{3.0}), let two vector functions}%
\begin{equation}
u_{1}\left( x,t\right) ,u_{2}\left( x,t\right) \in C^{2}\left( \overline{%
\Omega }_{h,T}\right)  \label{6.270}
\end{equation}%
\emph{\ are solutions of equation (\ref{3.1}) in }$\Omega _{h,T}$\emph{\ and 
}$\Gamma _{h,T}$ \emph{in (\ref{3.4}) is replaced with }$S_{h,T}.$\emph{\
Assume that conditions (\ref{6.2}) hold, in which }$G_{h,T}$\emph{\ is
replaced with }$\Omega _{h,T}.$\emph{\ Keep notations (\ref{6.03}). Then the
following Lipschitz stability estimate is valid:}%
\begin{equation}
\left\Vert \overline{u}\right\Vert _{H^{1}\left( \Omega _{h,T}\right) }\leq
C_{2}\left( \left\Vert \overline{g}_{0}\right\Vert _{H^{1}\left(
S_{h,T}\right) }+\left\Vert \overline{g}_{1}\right\Vert _{L_{2}\left(
S_{h,T}\right) }\right) ,  \label{6.28}
\end{equation}%
\emph{where norms in }$H^{1}\left( S_{h,T}\right) ,$ $L_{2}\left(
S_{h,T}\right) $\emph{\ are understood as in (\ref{2.03})-(\ref{2.310}). The
number }$C_{2}=C_{2}\left( M_{1},M_{2},\mu ,\Omega ,T,h_{0}\right) >0$\emph{%
\ depends only on listed parameters. In addition, problem }

\textbf{Proof.} By (\ref{6.270}), (\ref{2.31}) and (\ref{2.310}) norms in
the right hand side of (\ref{6.28}) make sense. Below $C_{2}>0$\ denotes
different numbers depending only on the above listed parameters.\emph{\ }%
Similarly with the proof of Theorem 6.1, we obtain the following analog of
inequality (\ref{6.10}):%
\begin{equation}
\left\vert L^{\left( s\right) }\left( \overline{u}\left( x,t_{s}\right)
\right) \right\vert \leq C_{2}\dsum\limits_{j=0}^{k}\left( \left\vert \nabla 
\overline{u}\left( x,t_{j}\right) \right\vert +\left\vert \overline{u}\left(
x,t_{j}\right) \right\vert \right) ,\text{ }x\in \Omega ;\text{ }%
t_{j},t_{s}\in Y.  \label{6.29}
\end{equation}%
In addition, 
\begin{equation}
\overline{u}\mid _{S_{h,T}}=\overline{g}_{0},\text{ }\partial _{l}\overline{u%
}\mid _{S_{h,T}}=\overline{g}_{1}.  \label{6.30}
\end{equation}%
Square both sides of each inequality (\ref{6.29}), apply Cauchy-Schwarz
inequality, then multiply by the function $\varphi _{\lambda ,\nu
_{0}}^{2}\left( x\right) ,$ where $\nu _{0}$ is the number of Theorem 5.1,
then integrate over the domain $\Omega $ and apply Carleman estimate (\ref%
{5.10}). In doing the latter, use Gauss formula to $\func{div}U_{1}$ and use
the estimate for $\left\vert U_{1}\right\vert $ in the second line of (\ref%
{5.10}). Then sum up resulting inequalities with respect to $s=0,...,k$.
Using (\ref{6.140}) we obtain for all $\lambda \geq \lambda _{0}$%
\begin{equation}
\left. 
\begin{array}{c}
\exp \left( 2\lambda \cdot 4^{\nu _{0}}\right) \left( \left\Vert \overline{g}%
_{0}\right\Vert _{H^{1}\left( S_{h,T}\right) }^{2}+\left\Vert \overline{g}%
_{1}\right\Vert _{L_{2}\left( S_{h,T}\right) }^{2}\right) + \\ 
+\dint\limits_{\Omega }\dsum\limits_{j=0}^{k}\left( \left\vert \nabla 
\overline{u}\left( x,t_{j}\right) \right\vert ^{2}+\left\vert \overline{u}%
\left( x,t_{j}\right) \right\vert ^{2}\right) \varphi _{\lambda ,\nu
_{0}}\left( x\right) dx\geq \\ 
\geq C_{2}\lambda \dint\limits_{\Omega }\dsum\limits_{j=0}^{k}\left(
\left\vert \nabla \overline{u}\left( x,t_{j}\right) \right\vert ^{2}+\lambda
^{2}\left\vert \overline{u}\left( x,t_{j}\right) \right\vert ^{2}\right)
\varphi _{\lambda ,\nu _{0}}\left( x\right) dx,\text{ }\forall \lambda \geq
\lambda _{0}.%
\end{array}%
\right.  \label{6.31}
\end{equation}%
Choose $\lambda _{2}=\lambda _{2}\left( M_{1},M_{2},\mu ,\Omega
,T,h_{0}\right) \geq \lambda _{0}$ depending only on listed parameters such
that $C_{2}\lambda _{2}/2>1.$ Hence, using (\ref{6.31}), we obtain 
\begin{equation}
\left. 
\begin{array}{c}
\exp \left( 2\cdot 4^{\nu _{0}}\cdot \lambda \right) \left( \left\Vert 
\overline{g}_{0}\right\Vert _{H^{1}\left( S_{h,T}\right) }^{2}+\left\Vert 
\overline{g}_{1}\right\Vert _{L_{2}\left( S_{h,T}\right) }^{2}\right) \geq
\\ 
\geq C_{2}\dint\limits_{\Omega }\dsum\limits_{j=0}^{k}\left( \left\vert
\nabla \overline{u}\left( x,t_{j}\right) \right\vert ^{2}+\left\vert 
\overline{u}\left( x,t_{j}\right) \right\vert ^{2}\right) \varphi _{\lambda
,\nu _{0}}^{2}\left( x\right) dx,\text{ }\forall \lambda \geq \lambda _{2}.%
\end{array}%
\right.  \label{6.32}
\end{equation}%
Next, it follows from (\ref{5.3}), (\ref{5.5}) and (\ref{5.6}) that $\varphi
_{\lambda ,\nu _{0}}^{2}\left( x\right) \geq \exp \left( 2\lambda \left(
4/5\right) ^{\nu _{0}}\right) $ in $\Omega .$ Hence, (\ref{6.32}) implies%
\begin{equation*}
C_{2}\left\Vert \overline{u}\right\Vert _{H^{1}\left( \Omega _{h,T}\right)
}\leq \exp \left[ 2\lambda \left( 4^{\nu _{0}}-\left( 4/5\right) ^{\nu
_{0}}\right) \right] \left( \left\Vert \overline{g}_{0}\right\Vert
_{H^{1}\left( S_{h,T}\right) }^{2}+\left\Vert \overline{g}_{1}\right\Vert
_{L_{2}\left( S_{h,T}\right) }^{2}\right) ,\text{ }\forall \lambda \geq
\lambda _{2}.
\end{equation*}%
Setting here $\lambda =\lambda _{2}=\lambda _{2}\left( M_{1},M_{2},\mu
,\Omega ,T,h_{0}\right) $ and recalling Remark 6.2, we obtain the target
estimate (\ref{6.28}). \ $\square $

Theorem 6.3 ensures the Lipschitz stability estimate (\ref{6.28}) in the $%
H^{1}\left( \Omega _{h,T}\right) -$ norm. We prove in Theorem 6.4 a similar
estimate in the stronger $H^{2}\left( \Omega _{h,T}\right) -$norm. Similarly
with Theorem 6.2, we should use in this case Carleman estimate (\ref{5.11})
instead of (\ref{5.10}). This is because (\ref{5.11}) contains derivatives
of the second order.

\textbf{Theorem 6.4 }(the second Lipschitz stability estimate). \emph{Assume
that conditions of Theorem 6.3 hold.\ Suppose that boundary functions}%
\begin{equation}
\emph{\ }\overline{g}_{0},\overline{g}_{1}\in H^{2}\left( S_{h,T}\right)
\label{6.033}
\end{equation}
\emph{(see Remarks 2.1) also that there exists a vector function }$p\left(
x,t\right) \in H^{2}\left( \Omega _{h,T}\right) $\emph{\ such that}%
\begin{equation}
\left. 
\begin{array}{c}
p\mid _{S_{h,T}}=\overline{g}_{0},\text{ }\partial _{l}p\mid _{S_{h,T}}=%
\overline{g}_{1}, \\ 
\left\Vert p\right\Vert _{H^{2}\left( \Omega _{h,T}\right) }\leq B_{1}\left(
\left\Vert \overline{g}_{0}\right\Vert _{H^{2}\left( S_{h,T}\right)
}+\left\Vert \overline{g}_{1}\right\Vert _{H^{2}\left( S_{h,T}\right)
}\right) ,%
\end{array}%
\right.  \label{6.33}
\end{equation}%
\emph{where the number }$B_{1}>0$\emph{\ is independent on functions }$%
\overline{g}_{0},\overline{g}_{1},$\emph{\ and norms in (\ref{6.33}) are
understood as in (\ref{2.03})-(\ref{2.310}). Keep notations (\ref{6.03}).
Then the following Lipschitz stability estimate holds:} 
\begin{equation}
\left\Vert \overline{u}\right\Vert _{H^{2}\left( \Omega _{h,T}\right) }\leq
C_{3}\left( \left\Vert \overline{g}_{0}\right\Vert _{H^{2}\left(
S_{h,T}\right) }+\left\Vert \overline{g}_{1}\right\Vert _{H^{2}\left(
S_{h,T}\right) }\right) ,  \label{6.34}
\end{equation}%
\emph{where the number }$C_{3}=C_{3}\left( M_{1},M_{2},B_{1},\mu ,\Omega
,T,h_{0}\right) >0$\emph{\ depends only on listed parameters.}

\textbf{Proof}. In this proof $C_{3}$ denotes different positive numbers
depending on the above parameters. By (\ref{2.31}), (\ref{2.310}) and (\ref%
{6.033}) norms in the right hand sides of (\ref{6.33}) and (\ref{6.34}) make
sense. Using the first line of (\ref{6.33}), and similarly with (\ref{6.25})
consider the following function $\widetilde{u}:$%
\begin{equation}
\widetilde{u}\left( x,t_{s}\right) =\overline{u}\left( x,t_{s}\right)
-p\left( x,t_{s}\right) ,\left( x,t_{s}\right) \in \Omega _{h,T}.
\label{6.35}
\end{equation}%
Then the first line of (\ref{6.33}) implies:%
\begin{equation}
\widetilde{u}\mid _{S_{h,T}}=\partial _{l}\widetilde{u}\mid _{S_{h,T}}=0.
\label{6.36}
\end{equation}%
Substituting $\overline{u}=\widetilde{u}+p$ in the first four lines of (\ref%
{6.6}) and turning the resulting equation in the inequality, we obtain the
following analog of (\ref{6.27}) 
\begin{equation}
\left. 
\begin{array}{c}
\left\vert L^{\left( s\right) }\left( \widetilde{u}\left( x,t_{s}\right)
\right) \right\vert \leq C_{3}\dsum\limits_{j=0}^{k}\left( \left\vert \nabla 
\widetilde{u}\left( x,t_{j}\right) \right\vert +\left\vert \widetilde{u}%
\left( x,t_{j}\right) \right\vert \right) + \\ 
+C_{3}\dsum\limits_{\left\vert \alpha \right\vert \leq
2}\dsum\limits_{j=0}^{k}\left\vert D_{x}^{\alpha }p\left( x,t_{j}\right)
\right\vert ,\text{ }x\in \Omega ;\text{ }t_{j},t_{s}\in Y.%
\end{array}%
\right.  \label{6.37}
\end{equation}%
Square both sides of each inequality (\ref{6.37}), apply Cauchy-Schwarz
inequality, then multiply by the CWF $\varphi _{\lambda ,\nu _{0}}\left(
x\right) ,$ apply Carleman estimate (\ref{5.11}) and then integrate the
resulting inequality over the domain $\Omega $, using Gauss formula. It
follows from the third and fourth lines of (\ref{5.11}) and (\ref{6.36})
that boundary integrals will not occur. Summing up resulting estimates with
respect to $s=0,...,k,$ using the second line of (\ref{6.33}) and (\ref{6.35}%
) and proceeding similarly with the part of the proof of Theorem 6.3, which
starts from (\ref{6.31}), we obtain the target estimate (\ref{6.34}). \ $%
\square $

\section{Stability Estimates for CIP1 in the TFD Framework (\protect\ref{4.9}%
)-(\protect\ref{4.12})}

\label{sec:7}

Theorems of this section address, within the TFD framework, the second long
standing question formulated in Introduction. We do not prove Theorems
7.1-7.4 since the TFD framework (\ref{4.9})-(\ref{4.12}) is just a linear
case of UCP (\ref{3.1})-(\ref{3.4}). Following (\ref{4.10}) and (\ref{4.12}%
), denote 
\begin{equation}
q_{0}\left( x,t\right) =\partial _{h,t}\widetilde{p}_{0}\left( x,t\right) ,%
\text{ }q_{1}\left( x,t\right) =\partial _{h,t}\widetilde{p}_{1}\left(
x,t\right) .  \label{7.1}
\end{equation}%
Then 
\begin{equation}
w\mid _{\Gamma _{h,T}}=q_{0}\left( x,t\right) ,\text{ }\partial _{l}w\mid
_{\Gamma _{h,T}}=q_{1}\left( x,t\right) \text{ in the case (\ref{4.10}),}
\label{7.2}
\end{equation}%
\begin{equation}
w\mid _{S_{h,T}}=q_{0}\left( x,t\right) ,\text{ }\partial _{l}w\mid
_{S_{h,T}}=q_{1}\left( x,t\right) \text{ in the case (\ref{4.12}).}
\label{7.3}
\end{equation}

Theorem 7.1 is an analog of Theorem 6.1. Recall Remark 5.1.

\textbf{Theorem 7.1. }\emph{Assume that in conditions (\ref{3.5}), (\ref%
{3.90}), (\ref{4.9})-(\ref{4.11}), the domain }$\Omega $\emph{\ is replaced
with the domain }$G$ of \emph{(\ref{5.7}),} \emph{and in so replaced
conditions (\ref{4.9})-(\ref{4.11}), the vector function }$w\in H^{2}\left(
G_{h,T}\right) .$\emph{\ Let conditions (\ref{1.1}), (\ref{7.1}) and (\ref%
{7.2}) hold. Suppose that conditions (\ref{3.50}) are replaced with}%
\begin{equation}
\left. 
\begin{array}{c}
a_{\alpha },\partial _{t}a_{\alpha }\in C\left( \overline{G}_{h,T}\right) 
\text{ for }\left\vert \alpha \right\vert \leq 1, \\ 
\left\Vert a_{\alpha }\right\Vert _{C\left( \overline{G}_{h,T}\right) }\leq
D,\text{ }\left\Vert \partial _{h,t}a_{\alpha }\right\Vert _{C\left( 
\overline{G}_{h,T}\right) }\leq D\text{ for }\left\vert \alpha \right\vert
\leq 1.%
\end{array}%
\right.  \label{7.4}
\end{equation}%
\emph{Let }$\delta \in \left( 0,1\right) $ \emph{be a number. Assume that
the following analog of (\ref{6.3}) is valid: }%
\begin{equation}
\left\Vert q_{0}\right\Vert _{H^{1}\left( \Gamma _{h,T}\right) }<\delta 
\text{ and }\left\Vert q_{1}\right\Vert _{L_{2}\left( \Gamma _{h,T}\right)
}<\delta .  \label{7.5}
\end{equation}%
\emph{Choose an arbitrary number }$\varepsilon \in \left( 0,1/6\right) .$%
\emph{\ Then there exists a sufficiently small number }$\delta _{3}=\delta
_{3}\left( a^{0},D,\mu ,G,T,c_{1},h_{0},\varepsilon \right) \in \left(
0,1\right) $\emph{\ such that the following H\"{o}lder stability estimates
hold:}%
\begin{equation}
\left\Vert w\right\Vert _{H^{1}\left( G_{3\varepsilon ,h,T}\right) }\leq
C_{4}\left( 1+\left\Vert w\right\Vert _{H^{1}\left( G_{h,T}\right) }\right)
\delta ^{\rho _{3}},\text{ }\forall \delta \in \left( 0,\delta _{3}\right) ,
\label{7.6}
\end{equation}%
\begin{equation}
\left\Vert \widetilde{a}_{\alpha _{0}}\right\Vert _{L_{2}\left(
G_{3\varepsilon ,h,T}\right) }\leq C_{4}\left( 1+\left\Vert w\right\Vert
_{H^{1}\left( G_{h,T}\right) }\right) \delta ^{\rho _{3}},\text{ }\forall
\delta \in \left( 0,\delta _{3}\right) ,  \label{7.7}
\end{equation}%
\emph{where the number }$\rho _{3}=\rho _{3}\left( a^{0},D,\mu
,G,T,c_{1},h_{0},\varepsilon \right) \in \left( 0,1\right) $\emph{\ and the
number }$C_{4}=C_{4}\left( a^{0},D,\mu ,G,T,c_{1},h_{0},\varepsilon \right)
>0.$\emph{\ Numbers }$\delta _{3},\rho _{3}$\emph{\ and }$C_{4}$\emph{\
depend only on listed parameters. In addition, problem (\ref{4.9})-(\ref%
{4.11}) has at most one solution }$\left( w,\widetilde{a}_{\alpha
_{0}}\right) \in H^{2}\left( G_{h,T}\right) \times L_{2}\left( G\right) .$ 
\emph{Also, there exists at most one vector function }$\left( w,\widetilde{a}%
_{\alpha _{0}}\right) \in H^{2}\left( G_{h,T}\right) \times L_{2}\left(
G\right) $\emph{\ conditions (\ref{4.9})-(\ref{4.11}).}

Theorem 7.2 is an analog of Theorem 6.2.

\textbf{Theorem 7.2}. \emph{Assume that all conditions of Theorem 7.1 hold,
except that estimates (\ref{7.5}) are replaced with estimates in stronger
norms, }%
\begin{equation*}
\left\Vert q_{0}\right\Vert _{H^{2}\left( \Gamma _{h,T}\right) }<\delta 
\text{ and }\left\Vert q_{1}\right\Vert _{H^{2}\left( \Gamma _{h,T}\right)
}<\delta ,
\end{equation*}%
\emph{see Remarks 2.1.} \emph{Then estimates (\ref{7.6}) and (\ref{7.7})
become:}%
\begin{equation*}
\left\Vert w\right\Vert _{H^{2}\left( G_{3\varepsilon ,h,T}\right) }\leq
C_{4}\left( 1+\left\Vert w\right\Vert _{H^{1}\left( G_{h,T}\right) }\right)
\delta ^{\rho _{4}},\text{ }\forall \delta \in \left( 0,\delta _{3}\right) ,
\end{equation*}%
\begin{equation*}
\left\Vert \widetilde{a}_{\alpha _{0}}\right\Vert _{L_{2}\left(
G_{3\varepsilon ,h,T}\right) }\leq C_{4}\left( 1+\left\Vert w\right\Vert
_{H^{1}\left( G_{h,T}\right) }\right) \delta ^{\rho _{4}},\text{ }\forall
\delta \in \left( 0,\delta _{3}\right) ,
\end{equation*}%
\emph{where the number }$\rho _{4}=\rho _{4}\left( a^{0},D,\mu
,G,T,c_{1},h_{0},\varepsilon \right) \in \left( 0,1\right) $ \emph{depends
only on listed parameters.}

Theorem 7.3 is an analog of Theorem 6.3.

\textbf{Theorem 7.3.} \emph{Assume that conditions (\ref{1.1}), (\ref{2.31}%
), (\ref{3.5}), (\ref{3.90}), (\ref{4.9}), (\ref{4.11}), (\ref{4.12}), (\ref%
{5.3}), (\ref{7.1}), (\ref{7.3}) and (\ref{7.4}) hold and the vector
function }$w\in H^{2}\left( \Omega _{h,T}\right) $\emph{. Then there exists
a number }$C_{5}=C_{5}\left( a^{0},D,\mu ,\Omega ,T,c_{1},h_{0}\right) >0$%
\emph{\ depending only on listed parameters such that the following
Lipschitz stability estimates hold:}%
\begin{equation}
\left\Vert w\right\Vert _{H^{1}\left( \Omega _{h,T}\right) }\leq C_{5}\left(
\left\Vert q_{0}\right\Vert _{H^{1}\left( S_{h,T}\right) }+\left\Vert
q_{1}\right\Vert _{L_{2}\left( S_{h,T}\right) }\right) ,  \label{7.8}
\end{equation}%
\begin{equation}
\left\Vert \widetilde{a}_{\alpha _{0}}\right\Vert _{L_{2}\left( \Omega
\right) }\leq C_{5}\left( \left\Vert q_{0}\right\Vert _{H^{1}\left(
S_{h,T}\right) }+\left\Vert q_{1}\right\Vert _{L_{2}\left( S_{h,T}\right)
}\right) .  \label{7.9}
\end{equation}%
\emph{In addition, problem (\ref{4.9}), (\ref{4.11}), (\ref{4.12}) has at
most one solution }$\left( w,\widetilde{a}_{\alpha _{0}}\right) \in
H^{2}\left( \Omega _{h,T}\right) \times L_{2}\left( \Omega \right) .$

Theorem 7.4 is an analog of Theorem 6.4.

\textbf{Theorem 7.4.} \emph{Assume that all conditions of Theorem 7.3 hold.
Suppose that functions }$q_{0},q_{1}\in H^{2}\left( S_{h,T}\right) $ \emph{%
(see Remarks 2.1). In addition, assume that there exists a vector function }$%
\widetilde{p}\left( x,t\right) \in H^{2}\left( \Omega _{h,T}\right) $\emph{\
such that}%
\begin{equation*}
\left. 
\begin{array}{c}
\widetilde{p}\mid _{S_{h,T}}=q_{0},\text{ }\partial _{l}\widetilde{p}\mid
_{S_{h,T}}=q_{1}, \\ 
\left\Vert \widetilde{p}\right\Vert _{H^{2}\left( \Omega _{h,T}\right) }\leq
B_{2}\left( \left\Vert q_{0}\right\Vert _{H^{2}\left( S_{h,T}\right)
}+\left\Vert q_{1}\right\Vert _{H^{2}\left( S_{h,T}\right) }\right) ,%
\end{array}%
\right.
\end{equation*}%
\emph{where the number }$B_{2}>0$\emph{\ is independent on functions }$%
q_{0},q_{1}.$ \emph{Then estimates} \emph{(\ref{7.8}) and (\ref{7.9}) become:%
}%
\begin{equation*}
\left. 
\begin{array}{c}
\left\Vert w\right\Vert _{H^{2}\left( \Omega _{h,T}\right) }\leq C_{6}\left(
\left\Vert q_{0}\right\Vert _{H^{2}\left( S_{h,T}\right) }+\left\Vert
q_{1}\right\Vert _{H^{2}\left( S_{h,T}\right) }\right) , \\ 
\left\Vert \widetilde{a}_{\alpha _{0}}\right\Vert _{L_{2}\left( \Omega
\right) }\leq C_{6}\left( \left\Vert q_{0}\right\Vert _{H^{2}\left(
S_{h,T}\right) }+\left\Vert q_{1}\right\Vert _{H^{2}\left( S_{h,T}\right)
}\right) ,%
\end{array}%
\right.
\end{equation*}%
\emph{where the number} $C_{6}=C_{6}\left( a^{0},D,\mu ,\Omega
,T,c_{1},B_{2},h_{0}\right) >0$\emph{\ depends only on listed parameters.}

\section{An Applied Example: Temporal and Spatial Monitoring of Epidemics}

\label{sec:8}

\subsection{The CIP of monitoring of epidemics}

\label{sec:8.1}

Let $\Omega \subset \mathbb{R}^{2}$ be a bounded domain modeling a city
where an epidemic occurs. Let $S$ be the number of susceptible patients, $I$
be the number of infected patients and $R$ be the number of recovered
patients in this city. Traditionally this so-called SIR model is governed by
a system of three coupled Ordinary Differential Equations (), which was
originally proposed by Kermack and McKendrick in 1927 \cite{Kermack}, also,
see \cite{Lee} for this system. However, this system describes only the
total number of SIR patients in the affected city at any moment of time. To
obtain both spatial and temporal distributions of SIR populations, Lee, Liu,
Tembine, Li and Osher \cite{Lee} have proposed \ a new model, which governs
hose distributions via a system of three coupled nonlinear parabolic PDEs.

It was noticed in \cite{epid} that coefficients of this system, which
describe infection and recovery rates, are unknown. Hence, a CIP of the
recovery of these coefficients was posed in \cite{epid} and a globally
convergent numerical method for it, the so-called convexification method,
was developed analytically and tested numerically in \cite{epid}, also, see
Introduction about the convexification method. The results of \cite{epid}
indicated that if those coefficients are recovered, then it is possible to
monitor spatial and temporal distributions of SIR\ populations inside of
that city using temporal measurements of both SIR populations and their
fluxes only at the boundary of the city. The latter is intriguing, since the
cost of monitoring might be significantly decreased.

However, it was assumed in the CIP of \cite{epid} that the lateral Cauchy \
data for that system are complemented by the measurements of its solution at 
$t=t_{0}\in \left( 0,T\right) $ inside the domain $\Omega .$ Stability
estimates for that CIP were not obtained in \cite{epid}, although uniqueness
theorem was proven. The goal of this section is to use the TFD framework to
prove Lipschitz stability estimate for the direct analog of CIP of \cite%
{epid}, assuming that $t_{0}=0.$ In the next section we develop a new
version of the convexification method for this CIP within the TFD framework.

\subsection{The CIP\ of monitoring epidemics within the TFD framework}

\label{sec:8.2}

We keep notations (\ref{1.2}), although, to shorten the presentation, we
consider only the case of complete data given at the entire lateral boundary 
$S_{T},$ thus setting $\Gamma =\varnothing .$ Let $u_{S}\left( x,t\right)
,u_{I}\left( x,t\right) $ and $u_{R}\left( x,t\right) $ be the numbers of
S,I and R populations respectively at the point $\left( x,t\right) \in
Q_{T}. $ The SIR system of parabolic PDEs is \cite[formulas (2.1)]{Lee}:%
\begin{equation}
\left. 
\begin{array}{c}
\partial _{t}u_{S}-\left( \eta _{S}^{2}/2\right) \Delta u_{S}+\func{div}%
\left( u_{S}q_{S}\right) +\beta \left( x\right) u_{S}u_{I}=0,\text{ in }%
Q_{T}, \\ 
\partial _{t}u_{I}-\left( \eta _{I}^{2}/2\right) \Delta u_{I}+\func{div}%
\left( u_{I}q_{I}\right) -\beta \left( x\right) u_{S}u_{I}+\gamma \left(
x\right) u_{I}=0,\text{ in }Q_{T}, \\ 
\partial _{t}u_{R}-\left( \eta _{R}^{2}/2\right) \Delta u_{R}+\func{div}%
\left( u_{R}q_{R}\right) -\gamma \left( x\right) u_{I}=0,\text{ in }Q_{T}.%
\end{array}%
\right.  \label{8.1}
\end{equation}%
Initial and boundary data for system (\ref{8.1}) are: 
\begin{equation}
u_{S}\left( x,0\right) =f_{S}\left( x\right) ,\text{ }u_{I}\left( x,0\right)
=f_{I}\left( x\right) ,\text{ }u_{R}\left( x,0\right) =f_{R}\left( x\right) ,%
\text{ in }\Omega ,  \label{8.2}
\end{equation}%
\begin{equation}
\partial _{l}u_{S}\mid _{S_{T}}=g_{S}\left( x,t\right) ,\ \partial
_{l}u_{I}\mid _{S_{T}}=g_{I}\left( x,t\right) ,\text{ }\partial
_{l}u_{R}\mid _{S_{T}}=g_{R}\left( x,t\right) .  \label{8.3}
\end{equation}%
In (\ref{8.1}), $\eta _{S}^{2},\eta _{R}^{2},\eta _{R}^{2}>0$ are the
so-called viscosity terms, $q_{S},q_{I}$ and $q_{R}$ are 2D vectors of
velocities of propagations of S,I and R populations respectively. We assume
solely for brevity that 
\begin{equation}
\frac{\eta _{S}^{2}}{2}\equiv \frac{\eta _{I}^{2}}{2}\equiv \frac{\eta
_{R}^{2}}{2}\equiv 1,  \label{8.4}
\end{equation}%
\begin{equation}
q_{S},q_{I},q_{R}\in \left( C^{1}\left( \overline{\Omega }\right) \right)
^{2}.  \label{8.5}
\end{equation}%
Normal derivatives in (\ref{8.3}) are fluxes of those populations through
the boundary of that city. Next, $\beta \left( x\right) $ and $\gamma \left(
x\right) $ are infection and recovery rates respectively.

System (\ref{8.1}) is a nonlinear one. Therefore, unlike the well
investigated linear case \cite[Chapter 4]{Lad}, existence of a sufficiently
smooth solution of problem (\ref{8.1})-(\ref{8.3}) can be proven only under
a certain assumption imposed on the length $T$ of the interval $\left(
0,T\right) ,$ see, e.g. \cite[Theorem 7.1 of Chapter 7]{Lad} for the case of
the Dirichlet boundary condition for one nonlinear parabolic PDE. Uniqueness
of the forward problem (\ref{8.1})-(\ref{8.3}) can also be proven by one of
classical techniques of \cite{Lad}. However, we are not interested in these
questions. Rather, we are interested in CIP2 which we pose below. Our
Lipschitz stability estimate of \ Theorem 8.1 immediately implies uniqueness
of that CIP within the TFD framework.

\textbf{Coefficient Inverse Problem 2 (CIP2).} \emph{Assume that there
exists the vector function }$\Phi \left( x,t\right) =\left(
u_{S},u_{I},u_{R}\right) \left( x,t\right) \in \left( C^{3}\left( \overline{%
\Omega }_{T}\right) \right) ^{3}$\emph{\ (see Remarks 2.1) satisfying
conditions (\ref{8.1})-(\ref{8.3}). Suppose that coefficients }$\beta \left(
x\right) ,\gamma \left( x\right) \in C\left( \overline{\Omega }\right) $%
\emph{\ are unknown. Also, assume that in addition to the right hand sides
of (\ref{8.2}), (\ref{8.3}), Dirichlet boundary conditions for functions }$%
u_{S},u_{I},u_{R}$\emph{\ are known,}%
\begin{equation}
u_{S}\mid _{S_{T}}=p_{S}\left( x,t\right) ,\ u_{I}\mid _{S_{T}}=p_{I}\left(
x,t\right) ,\text{ }u_{R}\mid _{S_{T}}=p_{R}\left( x,t\right) .  \label{8.6}
\end{equation}%
\emph{Find functions }$\beta \left( x\right) $\emph{\ and }$\gamma \left(
x\right) .$

We assume the availability of the data (\ref{8.6}) on the entire lateral
boundary $S_{T}$ for the sake of brevity only. In fact, results, similar
with ones of Theorems 6.1 and 6.2 for incomplete data, also hold true. The
knowledge of the right hand sides of (\ref{8.3}) and (\ref{8.6}) means that
both fluxes and the numbers of SIR populations are measured at the boundary
of the affected city. Our technique allows to find the vector function $\Phi
\left( x,t\right) $ along with the unknown coefficients $\beta \left(
x\right) ,\gamma \left( x\right) .$ This means that our technique allows one
to monitor spatial and temporal distributions of SIR populations inside of
the affected city using boundary measurements.

We now reformulate CIP2 in the TFD framework. Denote:%
\begin{equation}
\left. 
\begin{array}{c}
v_{1}\left( x,t\right) =\partial _{t}u_{S}\left( x,t\right) ,v_{2}\left(
x,t\right) =\partial _{t}u_{I}\left( x,t\right) ,v_{3}\left( x,t\right)
=\partial _{t}u_{R}\left( x,t\right) , \\ 
V\left( x,t\right) =\left( v_{1},v_{2},v_{3}\right) ^{T}\left( x,t\right) ,%
\text{ }V\in \left( C^{3}\left( \overline{\Omega }_{T}\right) \right) ^{3}.%
\end{array}%
\right.  \label{8.7}
\end{equation}%
By (\ref{8.2}) and (\ref{8.7})%
\begin{equation}
\left. 
\begin{array}{c}
u_{S}\left( x,t\right) =\dint\limits_{0}^{t}v_{1}\left( x,\tau \right) d\tau
+f_{S}\left( x\right) ,\text{ }u_{I}\left( x,t\right)
=\dint\limits_{0}^{t}v_{2}\left( x,\tau \right) d\tau +f_{I}\left( x\right) ,
\\ 
u_{R}\left( x,t\right) =\dint\limits_{0}^{t}v_{3}\left( x,\tau \right) d\tau
+f_{R}\left( x\right) ,%
\end{array}%
\right.  \label{8.8}
\end{equation}%
\begin{equation}
v_{i}\left( x,0\right) =v_{i}\left( x,t\right) -\dint\limits_{0}^{t}\partial
_{t}v_{i}\left( x,\tau \right) d\tau ,\text{ }i=1,2,3.  \label{8.9}
\end{equation}%
Assume that%
\begin{equation}
\left\vert f_{S}\left( x\right) \right\vert ,\text{ }\left\vert f_{I}\left(
x\right) \right\vert \geq c_{2}>0,  \label{8.10}
\end{equation}%
where $c_{2}>0$ is a positive number. Set $t=0$ in the first and third
equations (\ref{8.1}). Using (\ref{8.2}), (\ref{8.4}), (\ref{8.5}) (\ref{8.7}%
), (\ref{8.9}) and (\ref{8.10}), we obtain%
\begin{equation}
\beta \left( x\right) =\frac{1}{\left( f_{S}f_{I}\right) \left( x\right) }%
\left[ -v_{1}\left( x,t\right) +\dint\limits_{0}^{t}\partial _{t}v_{1}\left(
x,\tau \right) d\tau +\Delta f_{S}\left( x\right) -\func{div}\left(
f_{S}\left( x\right) q_{S}\left( x\right) \right) \right] ,  \label{8.11}
\end{equation}%
\begin{equation}
\gamma \left( x\right) =\frac{1}{f_{I}\left( x\right) }\left[ v_{3}\left(
x,t\right) -\dint\limits_{0}^{t}\partial _{t}v_{3}\left( x,\tau \right)
d\tau -\Delta f_{R}\left( x\right) +\func{div}\left( f_{R}\left( x\right)
q_{R}\left( x\right) \right) \right] .  \label{8.12}
\end{equation}

Differentiate equations (\ref{8.1}) with respect to $t$. Then, write the
resulting equations in finite differences with respect to $t$, using also
discrete versions (\ref{2.7}), (\ref{2.8}) of Volterra integrals and keeping
the same notations for semi-discrete versions $v_{1}\left( x,t_{s}\right)
,v_{2}\left( x,t_{s}\right) ,v_{3}\left( x,t_{s}\right) $ of functions $%
v_{1}\left( x,t\right) ,v_{2}\left( x,t\right) ,v_{3}\left( x,t\right) $,%
\begin{equation}
V\left( x,t_{s}\right) =\left( v_{1},v_{2},v_{3}\right) ^{T}\left(
x,t_{s}\right) \in \left( H^{2}\left( \Omega _{h,T}\right) \right) ^{3}.
\label{8.120}
\end{equation}%
The smoothness assumption in (\ref{8.120}) is reasonable due to the second
line of (\ref{8.7}) and (\ref{2.03}), also, see Remarks 2.1. In addition, we
use (\ref{8.4}), (\ref{8.5}) and (\ref{8.7})-(\ref{8.12}). We obtain three
integral differential equations. The first equation is:%
\begin{equation}
\left. 
\begin{array}{c}
A_{1}\left( V\right) \left( x,t_{s}\right) =\partial _{h,t}v_{1}\left(
x,t_{s}\right) -\Delta v_{1}\left( x,t_{s}\right) +\func{div}\left(
v_{1}\left( x,t_{s}\right) q_{S}\left( x\right) \right) + \\ 
+\left[ \left( f_{S}f_{I}\right) \left( x\right) \right] ^{-1}\times \\ 
\times \left[ -v_{1}\left( x,t_{s}\right) +\left(
\dint\limits_{0}^{t_{s}}\partial _{h,t}v_{1}\left( x,\tau \right) d\tau
\right) _{h}+\Delta f_{S}\left( x\right) -\func{div}\left( f_{S}\left(
x\right) q_{S}\left( x\right) \right) \right] \times \\ 
\times \left[ v_{1}\left( \left( \dint\limits_{0}^{t_{s}}v_{2}\left( x,\tau
\right) d\tau \right) _{h}+f_{I}\left( x\right) \right) +v_{2}\left( \left(
\dint\limits_{0}^{t_{s}}v_{1}\left( x,\tau \right) d\tau \right)
_{h}+f_{S}\left( x\right) \right) \right] = \\ 
=0\text{, }x\in \Omega ,t_{s}\in Y.%
\end{array}%
\right.  \label{8.13}
\end{equation}%
The second equation is:%
\begin{equation}
\left. 
\begin{array}{c}
A_{2}\left( V\right) \left( x,t_{s}\right) =\partial _{h,t}v_{2}\left(
x,t_{s}\right) -\Delta v_{2}\left( x,t_{s}\right) +\func{div}\left(
v_{2}\left( x,t_{s}\right) q_{I}\left( x\right) \right) - \\ 
-\left[ \left( f_{S}f_{I}\right) \left( x\right) \right] ^{-1}\times \\ 
\times \left[ -v_{1}\left( x,t_{s}\right) +\left(
\dint\limits_{0}^{t_{s}}\partial _{h,t}v_{1}\left( x,\tau \right) d\tau
\right) _{h}+\Delta f_{S}\left( x\right) -\func{div}\left( f_{S}\left(
x\right) q_{S}\left( x\right) \right) \right] \times \\ 
\times \left[ 
\begin{array}{c}
v_{1}\left( x,t_{s}\right) \left( \left( \dint\limits_{0}^{t_{s}}v_{2}\left(
x,\tau \right) d\tau \right) _{h}+f_{I}\left( x\right) \right) + \\ 
+v_{2}\left( x,t_{s}\right) \left( \left(
\dint\limits_{0}^{t_{s}}v_{1}\left( x,\tau \right) d\tau \right)
_{h}+f_{S}\left( x\right) \right)%
\end{array}%
\right] + \\ 
+\left( f_{I}\left( x\right) \right) ^{-1}\left[ 
\begin{array}{c}
v_{3}\left( x,t_{s}\right) -\left( \dint\limits_{0}^{t_{s}}\partial
_{h,t}v_{3}\left( x,\tau \right) d\tau \right) _{h}- \\ 
-\Delta f_{R}\left( x\right) +\func{div}\left( f_{R}\left( x\right)
q_{R}\left( x\right) \right)%
\end{array}%
\right] = \\ 
=0,\text{ }x\in \Omega ,t_{s}\in Y.%
\end{array}%
\right.  \label{8.14}
\end{equation}%
The third equation is:%
\begin{equation}
\left. 
\begin{array}{c}
A_{3}\left( V\right) \left( x,t_{s}\right) =\partial _{h,t}v_{3}\left(
x,t_{s}\right) -\Delta v_{3}\left( x,t_{s}\right) +\func{div}\left(
v_{3}\left( x,t_{s}\right) q_{R}\left( x\right) \right) - \\ 
-\left( f_{I}\left( x\right) \right) ^{-1}\left[ 
\begin{array}{c}
v_{3}\left( x,t_{s}\right) -\left( \dint\limits_{0}^{t_{s}}\partial
_{h,t}v_{3}\left( x,\tau \right) d\tau \right) _{h}- \\ 
-\Delta f_{R}\left( x\right) +\func{div}\left( f_{R}\left( x\right)
q_{R}\left( x\right) \right)%
\end{array}%
\right] v_{2}\left( x,t_{s}\right) = \\ 
=0,\text{ }x\in \Omega ,\text{ }t_{s}\in Y.\text{ }%
\end{array}%
\right.  \label{8.15}
\end{equation}%
Cauchy boundary data for system (\ref{8.13})-(\ref{8.15}) are obtained from (%
\ref{8.3}), (\ref{8.6}) and (\ref{8.7}),%
\begin{equation}
\left. 
\begin{array}{c}
Q=\left( v_{1},v_{2},v_{3}\right) \mid _{S_{h,T}}=\left( \partial
_{h,t}p_{S},\partial _{h,t}p_{I},\partial _{h,t}p_{R}\right) \left(
x,t_{s}\right) ,\text{ }t_{s}\in Y, \\ 
Z=\left( \partial _{l}v_{1},\partial _{l}v_{2},\partial _{l}v_{3}\right)
\mid _{S_{h,T}}=\left( \partial _{h,t}g_{S},g_{I},\partial
_{h,t}g_{R}\right) \left( x,t_{s}\right) ,\text{ }t_{s}\in Y.%
\end{array}%
\right.  \label{8.16}
\end{equation}%
Suppose that BVP (\ref{8.13})-(\ref{8.16}) is solved. Then, using (\ref{8.9}%
), (\ref{8.11}) and (\ref{8.12}), we set%
\begin{equation}
\left. 
\begin{array}{c}
\beta \left( x\right) =\left( \left( f_{S}f_{I}\right) \left( x\right)
\right) ^{-1}\left[ -v_{1}\left( x,0\right) +\Delta f_{S}\left( x\right) -%
\func{div}\left( f_{S}\left( x\right) q_{S}\left( x\right) \right) \right] ,
\\ 
\gamma \left( x\right) =\left( f_{I}\left( x\right) \right) ^{-1}\left[
v_{3}\left( x,0\right) -\Delta f_{R}\left( x\right) +\func{div}\left(
f_{R}\left( x\right) q_{R}\left( x\right) \right) \right] .%
\end{array}%
\right.  \label{8.17}
\end{equation}

\subsection{Lipschitz stability estimate for problem (\protect\ref{8.13})-(%
\protect\ref{8.17})}

\label{sec:8.3}

BVP (\ref{8.13})-(\ref{8.16}) is the full analog of UCP (\ref{3.0})-(\ref%
{3.4}), in which $\Gamma _{h,T}$ in (\ref{3.4}) should be replaced with $%
S_{h,T}.$ Hence, we will formulate an analog of Theorem 6.4. Recall that we
took in section 8 the complete data at $S_{h,T}$ instead of the incomplete
data at $\Gamma _{h,T}$ only for the sake of brevity. The second small
difference with problem (\ref{3.0})-(\ref{3.4}) is that we now have three
equations (\ref{8.13})-(\ref{8.15}) instead of one equation (\ref{3.1}). It
is clear, however, that Theorem 6.4 can easily be extended to this case. We
omit the proof of Theorem 8.1 since it is very similar with the proof of
Theorem 6.4.

First, we formulate analogs of conditions (\ref{6.1}), (\ref{6.2}) for our
specific case of problem (\ref{8.13})-(\ref{8.17}). Let $R>0$ be a number.
We assume that%
\begin{equation}
\left\Vert V\right\Vert _{\left( H^{2}\left( \Omega _{h,T}\right) \right)
^{3}}\leq R,  \label{8.19}
\end{equation}%
\begin{equation}
\left. 
\begin{array}{c}
\left\Vert q_{S}\right\Vert _{\left( C^{1}\left( \overline{\Omega }\right)
\right) ^{2}},\text{ }\left\Vert q_{I}\right\Vert _{\left( C^{1}\left( 
\overline{\Omega }\right) \right) ^{2}},\left\Vert q_{R}\right\Vert _{\left(
C^{1}\left( \overline{\Omega }\right) \right) ^{2}}\leq R, \\ 
\left\Vert f_{S}\right\Vert _{C^{2}\left( \overline{\Omega }\right) },\text{ 
}\left\Vert f_{I}\right\Vert _{C^{2}\left( \overline{\Omega }\right)
},\left\Vert f_{R}\right\Vert _{C^{2}\left( \overline{\Omega }\right) }\leq
R.%
\end{array}%
\right.  \label{8.20}
\end{equation}

Assume that there exist two vector functions $\left( V_{i}\left(
x,t_{s}\right) ,\beta _{i}\left( x\right) ,\gamma _{i}\left( x\right)
\right) ,$ $i=1,2$ satisfying conditions (\ref{8.120})-(\ref{8.19}) as well
as boundary conditions generated by (\ref{8.16}):%
\begin{equation}
\left. 
\begin{array}{c}
Q_{i}=V_{i}\mid _{S_{T}}=\left( \partial _{h,t}p_{i,S}\left( x,t_{s}\right)
,\partial _{h,t}p_{i,I},\partial _{h,t}p_{i,R}\right) \left( x,t_{s}\right) ,%
\text{ }t_{s}\in Y,\text{ }i=1,2, \\ 
Z_{i}=\partial _{l}V_{i}\mid _{S_{T}}=\left( \partial _{h,t}g_{i,S},\partial
_{h,t}g_{iI},\partial _{h,t}g_{i,R}\right) \left( x,t_{s}\right) ,\text{ }%
t_{s}\in Y,\text{ }i=1,2.%
\end{array}%
\right.  \label{8.21}
\end{equation}%
Denote%
\begin{equation}
\left. 
\begin{array}{c}
\widetilde{Q}\left( x,t_{s}\right) =\left( \partial _{h,t}\left(
p_{1,S}-p_{2,S}\right) ,\partial _{h,t}\left( p_{1,I}-p_{2,I}\right)
,\partial _{h,t}\left( p_{1,R}-p_{2,R}\right) \right) \left( x,t_{s}\right) ,%
\text{ } \\ 
\widetilde{Z}\left( x,t_{s}\right) =\left( \partial _{h,t}\left(
g_{1,S}-g_{2,S}\right) ,\partial _{h,t}\left( g_{1,I}-g_{2,I}\right)
,\partial _{h,t}\left( g_{1,R}-g_{2,R}\right) \right) \left( x,t_{s}\right) ,%
\text{ } \\ 
\widetilde{V}\left( x,t_{s}\right) =V_{1}\left( x,t_{s}\right) -V_{2}\left(
x,t_{s}\right) ,\text{ }x\in \Omega ,t_{s}\in Y.%
\end{array}%
\right.  \label{8.22}
\end{equation}%
Using (\ref{8.120}) and (\ref{8.21})-(\ref{8.22}), we obtain%
\begin{equation}
\widetilde{V}\mid _{S_{h,T}}=\widetilde{Q},\text{ }\partial _{l}\widetilde{V}%
\mid _{S_{h,T}}=\widetilde{Z}.  \label{8.25}
\end{equation}%
Following Remarks \ 2.1, we assume that 
\begin{equation}
\text{ }\widetilde{Q},\widetilde{Z}\in \left( H^{2}\left( S_{h,T}\right)
\right) ^{3}.  \label{8.250}
\end{equation}%
Suppose that there exists a vector function $W\left( x,t_{s}\right) $ such
that%
\begin{equation}
\left. 
\begin{array}{c}
W\in H^{2}\left( \Omega _{h,T}\right) ,\text{ }W\mid _{S_{h,T}}=\widetilde{Q}%
,\text{ }\partial _{l}W\mid _{S_{h,T}}=\widetilde{Z}, \\ 
\left\Vert W\right\Vert _{\left( H^{2}\left( \Omega _{h,T}\right) \right)
^{3}}\leq B_{3}\left( \left\Vert \text{ }\widetilde{Q}\right\Vert _{\left(
H^{2}\left( S_{h,T}\right) \right) ^{3}}+\left\Vert \text{ }\widetilde{Z}%
\right\Vert _{\left( H^{2}\left( S_{h,T}\right) \right) ^{3}}\right) ,%
\end{array}%
\right.  \label{8.26}
\end{equation}%
where the number $B_{3}>0$ is independent on vector functions $\widetilde{Q}$
and $\widetilde{Z}.$ In Theorem 8.1 we naturally mean the obvious 2D analog
of condition (\ref{5.3}).

\textbf{Theorem 8.1.}\emph{\ Let the initial conditions }$%
f_{S},f_{I},f_{R}\in C^{2}\left( \overline{\Omega }\right) .$ \emph{Assume
that conditions (\ref{1.1}), (\ref{2.31}), (\ref{5.3}), (\ref{8.10}) and (%
\ref{8.20}) hold. Assume that there exist two vector functions }$\left(
V_{i}\left( x,t_{s}\right) ,\beta _{i}\left( x\right) ,\gamma _{i}\left(
x\right) \right) ,$\emph{\ }$i=1,2$\emph{\ satisfying conditions (\ref{8.120}%
)-(\ref{8.19}) and (\ref{8.21})-(\ref{8.250}). Denote }$\widetilde{\beta }%
\left( x\right) =\beta _{1}\left( x\right) -\beta _{2}\left( x\right) ,$%
\emph{\ }$\widetilde{\gamma }\left( x\right) =\gamma _{1}\left( x\right)
-\gamma _{2}\left( x\right) .$\emph{\ In addition, assume that there exists
a vector function }$W$\emph{\ satisfying conditions (\ref{8.26}) with the
number }$B_{3}>0$\emph{\ independent on }$\widetilde{Q}$\emph{\ and }$%
\widetilde{Z}.$\emph{\ Then the following Lipschitz stability estimate is
valid:}%
\begin{equation*}
\left\Vert \widetilde{V}\right\Vert _{\left( H^{2}\left( \Omega
_{h,T}\right) \right) ^{3}}+\left\Vert \widetilde{\beta }\right\Vert
_{L_{2}\left( \Omega \right) }+\left\Vert \widetilde{\gamma }\right\Vert
_{L_{2}\left( \Omega \right) }\leq C_{7}\left( \left\Vert \text{ }\widetilde{%
Q}\right\Vert _{\left( H^{2}\left( S_{h,T}\right) \right) ^{3}}+\left\Vert 
\text{ }\widetilde{Z}\right\Vert _{\left( H^{2}\left( S_{h,T}\right) \right)
^{3}}\right) ,
\end{equation*}%
\emph{where the number }$C_{7}=C_{7}\left( \Omega
,R,T,c_{2},B_{3},h_{0}\right) >0\ $\emph{depends\ only\ on\ listed\
parameters. Problem has at most one solution }$\left( V,\beta ,\gamma
\right) \in \left( H^{2}\left( \Omega _{h,T}\right) \right) ^{3}\times
C\left( \overline{\Omega }\right) \times C\left( \overline{\Omega }\right) .$

\section{Convexification Method for Problem (\protect\ref{8.13})-(\protect
\ref{8.17})}

\label{sec:9}

\subsection{The convexification functional}

\label{sec:9.1}

To numerically solve the problem (\ref{8.13})-(\ref{8.17}), the
convexification method constructs a globally strongly cost functional for
finding the vector function $V\left( x,t_{s}\right) .$ We want to be close
to our previous numerical paper about this subject \cite{epid}. Recall that
in \cite{epid} a different version of the convexification was constructed
and numerically tested for the case $t_{0}\neq 0.$ Hence, it is convenient
now to replace the domain $\Omega $ in (\ref{5.3}) with the square%
\begin{equation}
\Omega =\left\{ x=\left( x_{1},x_{2}\right) :a<x_{1},x_{2}<b\right\} ,
\label{9.1}
\end{equation}%
where $0<a<b$ are some numbers, which is similar with formula (2.1) of \cite%
{epid}.

The next question is about the choice of the CWF. The CWF (\ref{5.5}), (\ref%
{5.6}) is good for the Carleman estimate for a general elliptic operator $%
L_{0}$ in (\ref{3.6}) with variable coefficients. However, due to its
dependence on two large parameters $\lambda ,\nu \geq 1,$ this function
changes too rapidly, which is inconvenient for computations. The same was
also observed in \cite{Baud1,Baud2,Baud3} for the second generation of the
convexification method for some CIPs for hyperbolic PDEs. Since the
principal part of each elliptic operator in equations (\ref{8.13})-(\ref%
{8.16}) is the Laplace operator, then a simpler CWF depending only on one
large parameter can work. Thus, keeping in mind a possible future numerical
implementation of the method introduced below, we use below a simpler CWF.
This \ CWF as well as the subspace $H_{0}^{2}\left( \Omega \right) \subset
H^{2}\left( \Omega \right) $ are: 
\begin{equation}
\left. 
\begin{array}{c}
\varphi _{\lambda }\left( x\right) =e^{2\lambda x_{1}^{2}}, \\ 
H_{0}^{2}\left( \Omega \right) =\left\{ u\in H^{2}\left( \Omega \right)
:u\mid _{\partial \Omega }=\partial _{l}u\mid _{\partial \Omega }=0\right\} .%
\end{array}%
\right.  \label{9.2}
\end{equation}

\textbf{Theorem 9.1 }(Carleman estimate \cite[Theorem 4.1]{KLZ19}, \cite[%
Theorem 8.4.1]{KL})\textbf{.} \emph{Let the domain }$\Omega $\emph{\ be as
in (\ref{9.1}) and the function }$\varphi _{\lambda }\left( x\right) $\emph{%
\ be as in (\ref{9.2}). Then there exists a sufficiently large number }$%
\lambda _{0}=\lambda _{0}\left( \Omega \right) \geq 1$\emph{\ and a number }$%
C=C\left( \Omega \right) >0,$\emph{\ both numbers depending only on }$\Omega
,$\emph{\ such that the following Carleman estimate holds: }%
\begin{equation}
\left. 
\begin{array}{c}
\dint\limits_{\Omega }\left( \Delta u\right) ^{2}\varphi _{\lambda }\left(
x\right) dx\geq \left( C/\lambda \right) \dint\limits_{\Omega }\left(
\dsum\limits_{i,j=1}^{2}u_{x_{i}x_{j}}^{2}\right) \varphi _{\lambda }\left(
x\right) dx+ \\ 
+C\lambda \dint\limits_{\Omega }\left( \left\vert \nabla u\right\vert
^{2}+u^{2}\right) \varphi _{\lambda }\left( x\right) dx,\text{ }\forall
\lambda \geq \lambda _{0},\text{ }\forall u\in H_{0}^{2}\left( \Omega
\right) .%
\end{array}%
\right.  \label{9.02}
\end{equation}

To avoid the use of the penalty regularization term, we now consider only
the case $V\in \left( H^{2}\left( \Omega _{h,T}\right) \right) ^{3}$ instead
of $V\in \left( C^{4}\left( \overline{\Omega }_{h,T}\right) \right) ^{3}$ in
(\ref{8.120}). Embedding theorem implies:%
\begin{equation}
\left\Vert V\right\Vert _{\left( C\left( \overline{\Omega }_{h,T}\right)
\right) ^{3}}\leq K\left\Vert V\right\Vert _{\left( H^{2}\left( \Omega
_{h,T}\right) \right) ^{3}},\text{ }\forall V\in \left( H^{2}\left( \Omega
_{h,T}\right) \right) ^{3},  \label{9.3}
\end{equation}%
where the number $K=K\left( \Omega ,h_{0},T\right) >0$ depends only on
listed parameters. Let $R>0$ be the number in (\ref{8.19}), (\ref{8.20}).
Introduce the set of functions $B\left( R\right) $ as:%
\begin{equation}
B\left( R\right) =\left\{ 
\begin{array}{c}
V\in \left( H^{2}\left( \Omega _{h,T}\right) \right) ^{3}:\text{ }\left\Vert
V\right\Vert _{\left( H^{2}\left( \Omega _{h,T}\right) \right) ^{3}}<R, \\ 
\text{ boundary conditions (\ref{8.16}) hold.}%
\end{array}%
\right\} .  \label{9.4}
\end{equation}

Let $A_{j}\left( V\right) \left( x,t_{s}\right) ,$ $j=1,2,3$ be the
nonlinear integral differential operators in the left hand sides of
equations (\ref{8.13})-(\ref{8.15}). We consider

\textbf{Minimization Problem}. \emph{Minimize the functional }$J_{\lambda
}\left( V\right) $ \emph{on the set }$\overline{B\left( R\right) },$\emph{\
where }%
\begin{equation}
J_{\lambda }\left( V\right) =\dint\limits_{\Omega }\left[ \dsum%
\limits_{j=1}^{3}\dsum\limits_{s=0}^{k}\left( A_{j}\left( V\right) \left(
x,t_{s}\right) \right) ^{2}\right] \varphi _{\lambda }\left( x\right) dx.
\label{9.5}
\end{equation}

Note that the penalty regularization term is not a part of this functional,
unlike our previous works on the convexification for CIPs for parabolic PDEs 
\cite{Kpar,epid}, \cite[section 9.3.2]{KL}. Also, theorems formulated below
claim that all required properties of $J_{\lambda }\left( V\right) $ take
place only for sufficiently large values of the parameter $\lambda .$ This
seems to be inconvenient since the function $\varphi _{\lambda }\left(
x\right) $ changes too rapidly then. Nevertheless, a rich computational
experience of all above cited publications on the convexification shows that
reasonable values $\lambda \in \left[ 1,5\right] $ are sufficient for
obtaining accurate reconstructions of unknown coefficients. This can be
explained by an analogy with any asymptotic theory. Indeed, such a theory
typically states that if a certain parameter $X$ is sufficiently large, then
a certain formula $Y$ is valid with a good accuracy. However, in specific
numerical studies only computational results can typically show which
exactly values of $X$ are sufficiently large to ensure a good accuracy of $Y$%
. And quite often these values are reasonable ones. More precisely, in \cite%
{epid} the optimal value $\lambda =3.$ On the other hand, Figures 1,2 of 
\cite{epid} indicate that if $\lambda =0$, i.e. in the case when the CWF $%
\varphi _{0}\left( x\right) \equiv 1,$ then the computational results might
likely be unsatisfactory.

\subsection{Theorems}

\label{sec:9.2}

Denote $\left[ ,\right] $ the scalar product in\emph{\ }$\left(
H_{0}^{2}\left( \Omega \right) \right) ^{3}.$

\textbf{Theorem 9.2.} \emph{Let the initial conditions }$f_{S},f_{I},f_{R}%
\in C^{2}\left( \overline{\Omega }\right) .$ \emph{Let (\ref{1.1}), (\ref%
{8.10}) and (\ref{8.20}) hold with certain constants }$h_{0},c_{2},M_{3}>0.$%
\emph{\ Let the domain }$\Omega $\emph{\ be as in (\ref{9.1}) and let }$%
\lambda _{0}\geq 1$\emph{\ be the number of Theorem 9.1. Then:}

\emph{1. At each point }$V\in \overline{B\left( R\right) },$\emph{\ the
functional }$J_{\lambda }\left( V\right) $\emph{\ has the Fr\'{e}chet
derivative }$J_{\lambda }^{\prime }\left( V\right) \in \left(
H_{0}^{2}\left( \Omega _{h,T}\right) \right) ^{3}.$\emph{\ Furthermore, this
derivative is Lipschitz continuous, i.e.}%
\begin{equation}
\left\Vert J_{\lambda }^{\prime }\left( V_{1}\right) -J_{\lambda }^{\prime
}\left( V_{2}\right) \right\Vert _{\left( H^{2}\left( \Omega _{h,T}\right)
\right) ^{3}}\leq M_{3}\left\Vert V_{1}-V_{2}\right\Vert _{\left(
H^{2}\left( \Omega _{h,T}\right) \right) ^{3}},\text{ }\forall
V_{1},V_{2}\in \overline{B\left( R\right) }  \label{9.50}
\end{equation}%
\emph{with a certain constant }$M_{3}=M_{3}\left( \lambda ,\Omega
,R,T,h_{0}\right) >0$\emph{\ depending only on listed parameters.}

\emph{2. There exists a sufficiently large number }$\lambda _{1}=\lambda
_{1}\left( \Omega ,R,T,c_{2},h_{0}\right) \geq \lambda _{0}$\emph{\ such
that the functional }$J_{\lambda }\left( V\right) $\emph{\ is strongly
convex on the set }$\overline{B\left( R\right) }$\emph{\ for all }$\lambda
\geq \lambda _{1},$\emph{\ i.e. the following inequality holds} 
\begin{equation}
\left. 
\begin{array}{c}
J_{\lambda }\left( V_{2}\right) -J_{\lambda }\left( V_{1}\right) -\left[
J_{\lambda }^{\prime }\left( V_{1}\right) ,V_{2}-V_{1}\right] \geq
C_{8}e^{2\lambda a^{2}}\left\Vert V_{2}-V_{1}\right\Vert _{\left(
H^{2}\left( \Omega _{h,T}\right) \right) ^{3}}^{2},\text{ } \\ 
\forall \lambda \geq \lambda _{1},\text{ }\forall V_{1},V_{2}\in \overline{%
B\left( R\right) },%
\end{array}%
\right.  \label{9.6}
\end{equation}%
\emph{where the number }$C_{8}=C_{8}\left( \Omega ,R,T,c_{2},h_{0}\right) >0$%
\emph{\ is independent on }$V_{1},V_{2}.$\emph{\ Both numbers }$\lambda _{1}$%
\emph{\ and }$C_{8}$\emph{\ depend only on listed parameters. }

\emph{3. For each }$\lambda \geq \lambda _{1}$\emph{\ there exists unique
minimizer }$V_{\lambda ,\min }\in \overline{B\left( R\right) }$\emph{\ of
the functional }$J_{\lambda }\left( V\right) $\emph{\ on the set }$\overline{%
B\left( R\right) }$\emph{\ and the following inequality holds:}%
\begin{equation}
\left[ J_{\lambda }^{\prime }\left( V_{\lambda ,\min }\right) ,V-V_{\lambda
,\min }\right] \geq 0,\text{ }\forall V\in \overline{B\left( R\right) }.
\label{9.7}
\end{equation}

Below $C_{8}=C_{8}\left( \Omega ,R,T,c_{2},h_{0}\right) >0$ denotes
different numbers depending on the listed parameters. In practice, the
boundary data (\ref{8.16}) are always given with noise of a level $\delta
\in \left( 0,1\right) .$ One of the main postulates of the regularization
theory is the assumption of the existence of the true solution for the
ideal, noiseless data \cite{T}. The vector function $V_{\lambda ,\min }$ is
called then the \textquotedblleft regularized solution". The question of the 
$\delta -$dependent estimations of the accuracy of the regularized solutions
is an important topic of the regularization theory. Theorem 9.3 addresses
this question for our case.

Let the true solution of problem (\ref{8.13})-(\ref{8.17}) be $\left(
V^{\ast },\beta ^{\ast },\gamma ^{\ast }\right) \in \left( H^{2}\left(
\Omega _{h,T}\right) \right) ^{3}\times C\left( \overline{\Omega }\right)
\times C\left( \overline{\Omega }\right) .$ The boundary data for $V^{\ast }$
are vector functions $Q^{\ast }$and $Z^{\ast },$ which are full analogs of
the boundary data (\ref{8.16}). Thus, 
\begin{equation}
V^{\ast }\mid _{S_{h,T}}=Q^{\ast },\text{ }\partial _{l}V^{\ast }\mid
_{S_{h,T}}=Z^{\ast }.  \label{9.8}
\end{equation}%
Let $B^{\ast }\left( R\right) $ be the full analog of the set $B\left(
R\right) $ in (\ref{9.4})$,$ in which, however, boundary conditions (\ref%
{8.16}) are replaced with right hand sides of (\ref{9.8}). Assume that there
exist vector functions $W,W^{\ast }$ such that 
\begin{equation}
\left. 
\begin{array}{c}
W\in B\left( R\right) ,W^{\ast }\in B^{\ast }\left( R\right) , \\ 
\left\Vert W-W^{\ast }\right\Vert _{\left( H^{2}\left( \Omega _{h,T}\right)
\right) ^{3}}<\delta .%
\end{array}%
\right.  \label{9.9}
\end{equation}

\textbf{Theorem 9.3 }(the accuracy of the regularized solution $V_{\lambda
,\min }$). \emph{Let }

$\lambda _{1}=\lambda _{1}\left( \Omega ,R,T,c_{2},h_{0}\right) \geq 1$\emph{%
\ be the number, which was found in Theorem 9.1. Assume that conditions (\ref%
{9.8}), (\ref{9.9}) hold. In the parameters for }$\lambda _{1}$\emph{\
replace }$R$\emph{\ with }$2R$\emph{\ and let }$\lambda _{2}$\emph{\ be such
that }$\lambda _{2}=\lambda _{1}\left( \Omega ,2R,T,c_{2},h_{0}\right) \geq
\lambda _{1}\left( \Omega ,R,T,c_{2},h_{0}\right) \geq 1.$\emph{\ For any }$%
\lambda \geq \lambda _{2}$ \emph{choose} \emph{the level of noise in the
data }$\delta \left( \lambda \right) $\emph{\ so small }%
\begin{equation}
\left( \exp \left( \lambda \left( b^{2}-a^{2}\right) \right) +1\right)
\delta \left( \lambda \right) =\widetilde{\delta }\left( \lambda \right) <R.
\label{9.10}
\end{equation}%
\emph{Let }%
\begin{equation}
V^{\ast }\in B^{\ast }\left( R-\widetilde{\delta }\left( \lambda \right)
\right) .  \label{9.11}
\end{equation}%
\emph{\ Fix a number }$\lambda \geq \lambda _{2}.$ \emph{Let }$V_{\lambda
,\min }\in \overline{B\left( R\right) }$\emph{\ be the unique minimizer of
the functional }$J_{\lambda }\left( V\right) $\emph{\ on this set, which was
found in Theorem 9.2. Then the following accuracy estimates hold:}%
\begin{equation}
\left. 
\begin{array}{c}
\left\Vert V_{\lambda ,\min }-V^{\ast }\right\Vert _{\left( H^{2}\left(
\Omega _{h,T}\right) \right) ^{3}}\leq \left( 1+\exp \left( \lambda \left(
b^{2}-a^{2}\right) \right) \right) \delta ,\text{ }\forall \delta \in \left(
0,\widetilde{\delta }\left( \lambda \right) \right) , \\ 
\left\Vert \beta _{\lambda ,\min }-\beta ^{\ast }\right\Vert _{L_{2}\left(
\Omega \right) }+\left\Vert \gamma _{\lambda ,\min }-\gamma ^{\ast
}\right\Vert _{L_{2}\left( \Omega \right) }\leq C_{8}\delta ,\text{ }\forall
\delta \in \left( 0,\widetilde{\delta }\left( \lambda \right) \right) ,%
\end{array}%
\right.  \label{9.12}
\end{equation}%
\emph{where functions }$\beta _{\lambda ,\min }\left( x\right) $\emph{\ and }%
$\gamma _{\lambda ,\min }\left( x\right) $\emph{\ are found from the first
and third components respectively of the vector function }$V_{\lambda ,\min
} $\emph{\ via full analogs of formulas (\ref{8.17}). }

Let $\lambda \geq \lambda _{2}$ and let the number $\widetilde{\delta }%
\left( \lambda \right) $ be the one defined in (\ref{9.10}). Let $\widetilde{%
\delta }\left( \lambda \right) $ be so small that%
\begin{equation}
\widetilde{\delta }\left( \lambda \right) \in \left( 0,\frac{R}{3}\right) .
\label{9.13}
\end{equation}
Let\emph{\ } 
\begin{equation}
V_{0}\in B\left( \frac{R}{3}-\widetilde{\delta }\left( \lambda \right)
\right)  \label{9.14}
\end{equation}%
be an arbitrary point. The gradient descent method of the minimization of
the functional $J_{\lambda }\left( V\right) $ constructs the following
sequence:%
\begin{equation}
V_{n}=V_{n-1}-\omega J_{\lambda }^{\prime }\left( V_{n-1}\right) ,\text{ }%
n=1,2,...,  \label{9.15}
\end{equation}%
where the step size $\omega >0.$ Note that since by Theorem 9.1 $J_{\lambda
}^{\prime }\left( V_{n-1}\right) \in \left( H_{0}^{2}\left( \Omega
_{h,T}\right) \right) ^{3},$ then the second line of (\ref{9.2}) implies
that all iterates $V_{n}$ have the same boundary conditions (\ref{8.16}).

\textbf{Theorem 9.4} (global convergence of the method (\ref{9.15})). \emph{%
Let }$\lambda \geq \lambda _{2}.$ \emph{Assume that conditions (\ref{9.8})-(%
\ref{9.10}), (\ref{9.13}) and (\ref{9.14}) hold. Let (\ref{9.11}) be
replaced with }%
\begin{equation}
V^{\ast }\in B^{\ast }\left( \frac{R}{3}-\widetilde{\delta }\left( \lambda
\right) \right) .  \label{9.16}
\end{equation}%
\emph{Then there exists a sufficiently small number }$\omega _{0}\in \left(
0,1\right) $\emph{\ such that for any }$\omega \in \left( 0,\omega
_{0}\right) $ there exists \emph{a number }$\theta =\theta \left( \omega
\right) \in \left( 0,1\right) $\emph{\ such that all iterates }$V_{n}\in
B\left( R\right) $\emph{\ and the following convergence estimate for the
sequence (\ref{9.15}) is valid:}%
\begin{equation}
\left. 
\begin{array}{c}
\left\Vert V_{n}-V^{\ast }\right\Vert _{\left( H^{2}\left( \Omega
_{h,T}\right) \right) ^{3}}+\left\Vert \beta _{n}-\beta ^{\ast }\right\Vert
_{L_{2}\left( \Omega \right) }+\left\Vert \gamma _{n}-\gamma ^{\ast
}\right\Vert _{L_{2}\left( \Omega \right) }\leq \\ 
\leq \left( 1+\exp \left( \lambda \left( b^{2}-a^{2}\right) \right) \right)
\delta +C_{8}\theta ^{n}\left\Vert V_{\lambda ,\min }-V_{0}\right\Vert
_{\left( H^{2}\left( \Omega _{h,T}\right) \right) ^{3}},\text{ }\forall
\delta \in \left( 0,\widetilde{\delta }\left( \lambda \right) \right) .%
\end{array}%
\right.  \label{9.17}
\end{equation}

\textbf{Remark 9.1.} \emph{We call a numerical method for a CIP globally
convergent, if there is a theorem, which claims that it delivers at least
one point in a sufficiently small neighborhood of the true solution without
any advanced knowledge of this neighborhood. Thus, since }$R>0$\emph{\ is an
arbitrary number and since }$V_{0}$\emph{\ is an arbitrary point in (\ref%
{9.14}), then (\ref{9.16}) and (\ref{9.17}) implies the global convergence
of the sequence (\ref{9.15}).}

\subsection{Proof of Theorem 9.2}

\label{sec:9.3}

Let $V_{1},V_{2}\in \overline{B\left( R\right) }$ be two arbitrary vector
functions and 
\begin{equation}
\widehat{V}=V_{2}-V_{1}.  \label{9.021}
\end{equation}
Then triangle inequality, (\ref{9.2}) and (\ref{9.4}) imply: 
\begin{equation}
\widehat{V}\in \overline{B_{0}\left( 2R\right) }=\left\{ V\in \left(
H_{0}^{2}\left( \Omega _{h,T}\right) \right) ^{3}:\left\Vert V\right\Vert
\leq 2R\right\} .  \label{9.21}
\end{equation}%
It follows from (\ref{9.5}) that to prove (\ref{9.6}), \ we need to analyze
the differences 
\begin{equation}
\left( A_{j}\left( V_{2}\right) \right) ^{2}-\left( A_{j}\left( V_{1}\right)
\right) ^{2}=\left( A_{j}\left( V_{1}+\widehat{V}\right) \right) ^{2}-\left(
A_{j}\left( V_{1}\right) \right) ^{2},\text{ }j=1,2,3,  \label{9.22}
\end{equation}%
where the operators $A_{j}$ are defined in (\ref{8.13})-(\ref{8.15}). Since
the operator $A_{3}$ has the simplest form out of these three, then, to save
space, we consider only the case of $A_{3}.$ Two other cases are completely
similar. By (\ref{8.15})%
\begin{equation}
\left( A_{3}\left( V_{1}+\widehat{V}\right) \right) ^{2}-\left( A_{3}\left(
V_{1}\right) \right) ^{2}=A_{3,\text{lin}}\left( \widehat{V}\right) +A_{3,%
\text{nonlin}}\left( \widehat{V}\right) ,  \label{9.23}
\end{equation}%
where $A_{3,\text{lin}}\left( \widehat{V}\right) $ and $A_{3,\text{nonlin}%
}\left( \widehat{V}\right) $ are linear and nonlinear operators respectively
with respect to $\widehat{V}.$ Let $V_{1}=\left(
v_{1,1},v_{2,1},v_{3,1}\right) $ and $\widehat{V}=\left( \widehat{v}_{1},%
\widehat{v}_{2},\widehat{v}_{3}\right) .$ The explicit form of $A_{3,\text{%
lin}}$ is: 
\begin{equation}
\left. 
\begin{array}{c}
A_{3,\text{lin}}\left( \widehat{V}\right) \left( x,t_{s}\right) =2A_{3,\text{%
lin}}^{\left( 1\right) }\left( \widehat{V}\right) \left( x,t_{s}\right)
\cdot A_{3}\left( V_{1}\right) \left( x,t_{s}\right) , \\ 
A_{3,\text{lin}}^{\left( 1\right) }\left( \widehat{V}\right) \left(
x,t_{s}\right) =\left( \partial _{h,t}\widehat{v}_{3}\left( x,t_{s}\right)
-\Delta \widehat{v}_{3}\left( x,t_{s}\right) +\func{div}\left( \widehat{v}%
_{3}\left( x,t_{s}\right) q_{R}\left( x\right) \right) \right) - \\ 
-\left( f_{I}\left( x\right) \right) ^{-1}\left[ \widehat{v}_{3}\left(
x,t_{s}\right) -\left( \dint\limits_{0}^{t_{s}}\partial _{h,t}\widehat{v}%
_{3}\left( x,\tau \right) d\tau \right) _{h}\right] v_{2,1}\left(
x,t_{s}\right) - \\ 
-\left( f_{I}\left( x\right) \right) ^{-1}\left[ v_{3,1}\left(
x,t_{s}\right) -\left( \dint\limits_{0}^{t_{s}}\partial _{h,t}v_{3,1}\left(
x,\tau \right) d\tau \right) _{h}\right] \widehat{v}_{2}\left( x,t_{s}\right)
\\ 
-\left( f_{I}\left( x\right) \right) ^{-1}\left( -\Delta f_{R}\left(
x\right) +\func{div}\left( f_{R}\left( x\right) q_{R}\left( x\right) \right)
\right) \widehat{v}_{2}\left( x,t_{s}\right) ,\text{ }x\in \Omega ,\text{ }%
t_{s}\in Y.%
\end{array}%
\right.  \label{9.24}
\end{equation}%
The explicit form of $A_{3,\text{nonlin}}\left( \widehat{V}\right) \left(
x,t_{s}\right) $ is:%
\begin{equation}
A_{3,\text{nonlin}}\left( \widehat{V}\right) \left( x,t_{s}\right) =\left[
A_{3,\text{lin}}^{\left( 1\right) }\left( \widehat{V}\right) \right]
^{2}\left( x,t_{s}\right) ,\text{ }x\in \Omega ,\text{ }t_{s}\in Y.
\label{9.25}
\end{equation}%
Since formulas, similar with ones of (\ref{9.23})-(\ref{9.25}), are also
valid for $j=1,2$ in (\ref{9.22}), then (\ref{9.5}) leads to:%
\begin{equation}
\left. 
\begin{array}{c}
J_{\lambda }\left( V_{1}+\widehat{V}\right) -J_{\lambda }\left( V_{1}\right)
=\dint\limits_{\Omega }\left[ \dsum\limits_{j=1}^{3}\dsum%
\limits_{s=0}^{k}A_{j,\text{lin}}\left( \widehat{V}\right) \left(
x,t_{s}\right) \right] \varphi _{\lambda }\left( x\right) dx+ \\ 
+\dint\limits_{\Omega }\dsum\limits_{j=1}^{3}\dsum\limits_{s=0}^{k}\left[
A_{j,\text{lin}}^{\left( 1\right) }\left( \widehat{V}\right) \right]
^{2}\left( x,t_{s}\right) \varphi _{\lambda }\left( x\right) dx,%
\end{array}%
\right.  \label{9.26}
\end{equation}%
where $A_{j,\text{lin}}\left( \widehat{V}\right) \left( x,t_{s}\right) $ for 
$j=1,2$ depend linearly on $\widehat{V},$ and they are similar with $A_{3,%
\text{lin}}\left( \widehat{V}\right) \left( x,t_{s}\right) .$ Also, $\left[
A_{j,\text{lin}}^{\left( 1\right) }\left( \widehat{V}\right) \right]
^{2}\left( x\right) $ for $j=1,2$ are similar with (\ref{9.25}).

Since $\widehat{V}\in \left( H_{0}^{2}\left( \Omega _{h,T}\right) \right)
^{3}$ by (\ref{9.21}), then the integral term in the first line of (\ref%
{9.26}) is a bounded linear functional $\widetilde{J}_{\lambda }\left( 
\widehat{V}\right) :\left( H_{0}^{2}\left( \Omega _{h,T}\right) \right)
^{3}\rightarrow \mathbb{R}.$ Next it follows from (\ref{9.26}) that 
\begin{equation*}
\lim_{\left\Vert \widehat{V}\right\Vert _{\left( H^{2}\left( \Omega
_{h,T}\right) \right) ^{3}}}\frac{J_{\lambda }\left( V_{1}+\widehat{V}%
\right) -J_{\lambda }\left( V_{1}\right) -\widetilde{J}_{\lambda }\left( 
\widehat{V}\right) }{\left\Vert \widehat{V}\right\Vert _{\left( H^{2}\left(
\Omega _{h,T}\right) \right) ^{3}}}=0.
\end{equation*}%
Hence, $\widetilde{J}_{\lambda }\left( \widehat{V}\right) $ is the Fr\'{e}%
chet derivative of the functional $J_{\lambda }\left( V\right) $ at the
point $V_{1}.$ By Riesz theorem, there exists a point $J_{\lambda }^{\prime
}\left( V_{1}\right) \in \left( H_{0}^{2}\left( \Omega _{h,T}\right) \right)
^{3}$ such that $\widetilde{J}_{\lambda }\left( \widehat{V}\right) =\left[
J_{\lambda }^{\prime }\left( V_{1}\right) ,\widehat{V}\right] $ for all $%
\widehat{V}\in \left( H_{0}^{2}\left( \Omega _{h,T}\right) \right) ^{3}.$
Thus, the existence of the Fr\'{e}chet derivative of the functional $%
J_{\lambda }\left( V\right) $ is established. Its Lipschitz continuity
property (\ref{9.50}) can be proven similarly with \cite[Theorem 3.1]{Bak}
and \cite[Theorem 5.3.1]{KL}. Thus, we omit the proof of (\ref{9.50}).

Note that until now we have not used the Carleman estimate of Theorem 9.1.
Now, however, we focus on the proof of the strong convexity property (\ref%
{9.6}) and use, therefore, Theorem 9.1. Using (\ref{2.4})-(\ref{2.8}), (\ref%
{8.19}), Cauchy-Schwarz inequality, (\ref{9.24}), (\ref{9.25}) as well as
obvious analogs of the two latter formulas for $A_{j,\text{nonlin}}\left( 
\widehat{V}\right) \left( x,t_{s}\right) ,$ $j=1,2,$ we obtain%
\begin{equation}
\left. 
\begin{array}{c}
\left[ A_{j,\text{lin}}^{\left( 1\right) }\left( \widehat{V}\right) \right]
^{2}\left( x,t_{s}\right) \geq \left( 1/2\right) \left( \Delta \widehat{v}%
_{j}\right) ^{2}\left( x,t_{s}\right)
-C_{8}\dsum\limits_{m=1}^{3}\dsum\limits_{i=0}^{k}\left( \left\vert \nabla 
\widehat{v}_{m}\right\vert ^{2}+\widehat{v}_{m}^{2}\right) \left(
x,t_{i}\right) ,\text{ } \\ 
j=1,2,3;\text{ }x\in \Omega ;\text{ }t_{i},t_{s}\in Y.%
\end{array}%
\right.  \label{9.27}
\end{equation}
Hence, using (\ref{9.26}) and (\ref{9.27}), we obtain 
\begin{equation*}
\left. 
\begin{array}{c}
J_{\lambda }\left( V_{1}+\widehat{V}\right) -J_{\lambda }\left( V_{1}\right)
-\left[ J_{\lambda }^{\prime }\left( V_{1}\right) ,\widehat{V}\right] \geq
\\ 
\geq \left( 1/2\right) \dint\limits_{\Omega
}\dsum\limits_{j=1}^{3}\dsum\limits_{s=0}^{k}\left( \Delta \widehat{v}%
_{j}\right) ^{2}\left( x,t_{s}\right) \varphi _{\lambda }\left( x\right)
dx-C_{8}\dint\limits_{\Omega
}\dsum\limits_{j=1}^{3}\dsum\limits_{s=0}^{k}\left( \left\vert \nabla 
\widehat{v}_{j}\right\vert ^{2}+\widehat{v}_{j}^{2}\right) \left(
x,t_{s}\right) \varphi _{\lambda }\left( x\right) dx.%
\end{array}%
\right.
\end{equation*}%
Applying Carleman estimate (\ref{9.02}) to the second line of this
inequality, we obtain%
\begin{equation*}
\left. 
\begin{array}{c}
J_{\lambda }\left( V_{1}+\widehat{V}\right) -J_{\lambda }\left( V_{1}\right)
-\left[ J_{\lambda }^{\prime }\left( V_{1}\right) ,\widehat{V}\right] \geq
\\ 
\geq \left( C/\lambda \right) \dint\limits_{\Omega
}\dsum\limits_{m,r=1}^{2}\dsum\limits_{j=1}^{3}\dsum\limits_{s=0}^{k}\left(
\left( \widehat{v}_{j}\right) _{x_{m}x_{r}}\right) ^{2}\left( x,t_{s}\right)
\varphi _{\lambda }\left( x\right) dx+ \\ 
+C\lambda \dint\limits_{\Omega }\dsum\limits_{j=1}^{3}\dsum\limits_{s=0}^{k} 
\left[ \left( \nabla \widehat{v}_{j}\right) ^{2}\left( x,t_{s}\right)
+\lambda ^{2}\nabla \widehat{v}_{j}^{2}\left( x,t_{s}\right) \right] \varphi
_{\lambda }\left( x\right) dx- \\ 
-C_{8}\dint\limits_{\Omega
}\dsum\limits_{j=1}^{3}\dsum\limits_{s=0}^{k}\left( \left\vert \nabla 
\widehat{v}_{j}\right\vert ^{2}+\widehat{v}_{j}^{2}\right) \left(
x,t_{s}\right) \varphi _{\lambda }\left( x\right) dx,\text{ }\forall \lambda
\geq \lambda _{0}.%
\end{array}%
\right.
\end{equation*}%
Hence, there exists a sufficiently large number $\lambda _{1}=\lambda
_{1}\left( \Omega ,T,M_{3},c_{2},h_{0}\right) \geq \lambda _{0}$ such that 
\begin{equation}
\left. 
\begin{array}{c}
J_{\lambda }\left( V_{1}+\widehat{V}\right) -J_{\lambda }\left( V_{1}\right)
-\left[ J_{\lambda }^{\prime }\left( V_{1}\right) ,\widehat{V}\right] \geq
\\ 
\geq C_{8}\dint\limits_{\Omega }\left[ \dsum\limits_{j=1}^{3}\dsum%
\limits_{s=0}^{k}\left( \dsum\limits_{m,r=1}^{2}\left( \left( \widehat{v}%
_{j}\right) _{x_{m}x_{r}}\right) ^{2}+\left\vert \nabla \widehat{v}%
_{j}\right\vert ^{2}+\widehat{v}_{j}^{2}\right) \left( x,t_{s}\right) \right]
\varphi _{\lambda }\left( x\right) dx, \\ 
\forall \lambda \geq \lambda _{1}.%
\end{array}%
\right.  \label{9.28}
\end{equation}%
Since by (\ref{9.1}) and (\ref{9.2}) $\varphi _{\lambda }\left( x\right)
\geq e^{2\lambda a^{2}}$ in $\Omega ,$ then (\ref{9.28}), (\ref{2.03}) and (%
\ref{9.021}) imply (\ref{9.6}). Finally, we omit proofs of the existence and
uniqueness of the minimizer $V_{\lambda ,\min }\in \overline{B\left(
R\right) }$ as well as of inequality (\ref{9.7}) since these statements
follow from a combination of (\ref{9.50}) and (\ref{9.6}) with two results
of \cite{Bak}: Lemma 2.1 and Theorem 2.1. $\square $

\subsection{Proof of Theorem 9.3}

\label{sec:9.4}

Let $V\in \overline{B\left( R\right) }$ be an arbitrary vector function.
Denote 
\begin{equation}
\overline{V}=V-W,\text{ }\overline{V}^{\ast }=V^{\ast }-W^{\ast }.
\label{9.29}
\end{equation}%
Using the first line of (\ref{9.9}), (\ref{9.11}), (\ref{9.21}) and (\ref%
{9.29}), we obtain 
\begin{equation}
\overline{V},\overline{V}^{\ast }\in \overline{B_{0}\left( 2R\right) }.
\label{9.30}
\end{equation}
Consider the functional $I_{\lambda },$%
\begin{equation}
I_{\lambda }:\overline{B_{0}\left( 2R\right) }\rightarrow \mathbb{R},\text{ }%
I_{\lambda }\left( \overline{V}\right) =J_{\lambda }\left( \overline{V}%
+W\right) .  \label{9.31}
\end{equation}%
Then the full analog of Theorem 9.2 is valid for the functional $I_{\lambda
}\left( \overline{V}\right) $ for all $\lambda \geq \lambda _{2}.$ Let $%
\overline{V}_{\lambda ,\min }\in \overline{B_{0}\left( 2R\right) }$ be the
unique minimizer of $I_{\lambda }\left( \overline{V}\right) $ on the set $%
\overline{B_{0}\left( 2R\right) },$ which was found in that analog. Then by
the analog of (\ref{9.7}) 
\begin{equation}
\left[ I_{\lambda }^{\prime }\left( \overline{V}_{\lambda ,\min }\right) ,%
\overline{V}-\overline{V}_{\lambda ,\min }\right] \geq 0,\text{ }\forall 
\overline{V}\in \overline{B_{0}\left( 2R\right) },\text{ }\forall \lambda
\geq \lambda _{2}.  \label{9.32}
\end{equation}%
By the analog of (\ref{9.6}), (\ref{9.29}) and (\ref{9.30}) 
\begin{equation}
\left. 
\begin{array}{c}
I_{\lambda }\left( \overline{V}^{\ast }\right) -I_{\lambda }\left( \overline{%
V}_{\lambda ,\min }\right) -\left[ I_{\lambda }^{\prime }\left( \overline{V}%
_{\lambda ,\min }\right) ,\overline{V}^{\ast }-\overline{V}_{\lambda ,\min }%
\right] \geq \\ 
\geq C_{8}e^{2\lambda a^{2}}\left\Vert \overline{V}^{\ast }-\overline{V}%
_{\lambda ,\min }\right\Vert _{\left( H^{2}\left( \Omega _{h,T}\right)
\right) ^{3}}^{2},\text{ }\forall \lambda \geq \lambda _{2}.%
\end{array}%
\right.  \label{9.33}
\end{equation}%
Using (\ref{9.32}), we obtain $-\left[ I_{\lambda }^{\prime }\left( 
\overline{V}_{\lambda ,\min }\right) ,\overline{V}^{\ast }-\overline{V}%
_{\lambda ,\min }\right] \leq 0.$ Next, $-I_{\lambda }\left( \overline{V}%
_{\lambda ,\min }\right) \leq 0.$ Hence, (\ref{9.33}) implies%
\begin{equation}
\left\Vert \overline{V}^{\ast }-\overline{V}_{\lambda ,\min }\right\Vert
_{\left( H^{2}\left( \Omega _{h,T}\right) \right) ^{3}}\leq \frac{%
e^{-\lambda a^{2}}}{\sqrt{C_{8}}}\sqrt{I_{\lambda }\left( \overline{V}^{\ast
}\right) }.  \label{9.34}
\end{equation}

Estimate now $I_{\lambda }\left( \overline{V}^{\ast }\right) .$ By (\ref%
{8.13})-(\ref{8.15}), (\ref{9.1}), (\ref{9.2}), (\ref{9.9}), (\ref{9.29})
and (\ref{9.31})%
\begin{equation}
\left. 
\begin{array}{c}
I_{\lambda }\left( \overline{V}^{\ast }\right) =J_{\lambda }\left( \overline{%
V}^{\ast }+W\right) =J_{\lambda }\left( \left( \overline{V}^{\ast }+W^{\ast
}\right) +\left( W-W^{\ast }\right) \right) \leq \\ 
\leq 2J_{\lambda }\left( V^{\ast }\right) +C_{8}\delta ^{2}e^{2\lambda
b^{2}}.%
\end{array}%
\right.  \label{9.35}
\end{equation}%
However, $J_{\lambda }\left( V^{\ast }\right) =0.$ Hence, using (\ref{9.34})
and (\ref{9.35}), we obtain 
\begin{equation*}
\left\Vert \overline{V}^{\ast }-\overline{V}_{\lambda ,\min }\right\Vert
_{\left( H^{2}\left( \Omega _{h,T}\right) \right) ^{3}}\leq e^{\lambda
\left( b^{2}-a^{2}\right) }\delta ,\text{ }\forall \lambda \geq \lambda _{2}.
\end{equation*}%
Hence, using triangle inequality, (\ref{9.9}) and (\ref{9.29}), we obtain%
\begin{equation}
\left\Vert \left( \overline{V}_{\lambda ,\min }+W\right) -V^{\ast
}\right\Vert _{\left( H^{2}\left( \Omega _{h,T}\right) \right) ^{3}}\leq
\left( e^{\lambda \left( b^{2}-a^{2}\right) }+1\right) \delta ,\text{ }%
\forall \lambda \geq \lambda _{2}.  \label{9.37}
\end{equation}%
Let $\widetilde{\delta }\left( \lambda \right) $ be the number defined in (%
\ref{9.10}) and let $\delta \in \left( 0,\widetilde{\delta }\left( \lambda
\right) \right) .$ Then it follows from (\ref{9.11}), (\ref{9.37}) and
triangle inequality that 
\begin{equation}
\left( \overline{V}_{\lambda ,\min }+W\right) \in B\left( R\right) .
\label{9.38}
\end{equation}%
Let $V_{\lambda ,\min }\in \overline{B\left( R\right) }$ be the unique
minimizer of the functional $J_{\lambda }\left( V\right) $ on the set $%
\overline{B\left( R\right) },$ which was found in Theorem 9.2. Then by (\ref%
{9.31}) and (\ref{9.38}) 
\begin{equation*}
J_{\lambda }\left( V_{\lambda ,\min }\right) \leq J_{\lambda }\left( 
\overline{V}_{\lambda ,\min }+W\right) =I_{\lambda }\left( \overline{V}%
_{\lambda ,\min }\right) .
\end{equation*}%
This inequality is equivalent with 
\begin{equation}
J_{\lambda }\left( V_{\lambda ,\min }\right) =J_{\lambda }\left( \left(
V_{\lambda ,\min }-W\right) +W\right) =I_{\lambda }\left( V_{\lambda ,\min
}-W\right) \leq I_{\lambda }\left( \overline{V}_{\lambda ,\min }\right) .
\label{9.39}
\end{equation}%
Since $\left( V_{\lambda ,\min }-W\right) \in \overline{B_{0}\left(
2R\right) },$ then (\ref{9.39}) implies that $\left( V_{\lambda ,\min
}-W\right) $ is another minimizer, in addition to $\overline{V}_{\lambda
,\min },$ of the functional $I_{\lambda }\left( \overline{V}\right) $ on the
set $\overline{B_{0}\left( 2R\right) }.$ However, since such a minimizer is
unique, then $V_{\lambda ,\min }-W=\overline{V}_{\lambda ,\min }.$ Hence, $%
V_{\lambda ,\min }=\overline{V}_{\lambda ,\min }+W.$ This and (\ref{9.37})
apply the estimate in the first line of (\ref{9.12}) immediately. The second
estimate (\ref{9.12}) easily follows from a combination of the first one
with full analogs of formulas (\ref{8.17}). \ $\square $

\subsection{Proof of Theorem 9.4}

\label{sec:9.5}

Recall that $\lambda \geq \lambda _{2}$. Let again $V_{\lambda ,\min }\in 
\overline{B\left( R\right) }$\emph{\ }be the unique minimizer of the
functional $J_{\lambda }\left( V\right) $\ on this set, which was found in
Theorem 9.2.\emph{\ }By (\ref{9.10}), the first line of (\ref{9.12}), (\ref%
{9.16}) and triangle inequality 
\begin{equation*}
\left\Vert V_{\lambda ,\min }\right\Vert _{\left( H^{2}\left( \Omega
_{h,T}\right) \right) ^{3}}\leq \left\Vert V^{\ast }\right\Vert _{\left(
H^{2}\left( \Omega _{h,T}\right) \right) ^{3}}+\widetilde{\delta }\left(
\lambda \right) <\frac{R}{3},\text{ }\forall \delta \in \left( 0,\widetilde{%
\delta }\left( \lambda \right) \right) .
\end{equation*}%
Hence, by (\ref{9.4}) $V_{\lambda ,\min }\in B\left( R/3\right) .$ Hence,
applying Theorem 6 of \cite{SAR}, we obtain the existence of a sufficiently
small number $\omega _{0}\in \left( 0,1\right) $\ such that for any $\omega
\in \left( 0,\omega _{0}\right) $ there exists a number $\theta =\theta
\left( \omega \right) \in \left( 0,1\right) $\ such that $V_{n}\in B\left(
R\right) $ for all $n=0,1,...$\ and also%
\begin{equation}
\left\Vert V_{\lambda ,\min }-V_{n}\right\Vert _{\left( H^{2}\left( \Omega
_{h,T}\right) \right) ^{3}}\leq C_{8}\theta ^{n}\left\Vert V_{\lambda ,\min
}-V_{0}\right\Vert _{\left( H^{2}\left( \Omega _{h,T}\right) \right) ^{3}},%
\text{ }n=1,...,  \label{9.40}
\end{equation}%
for all $\delta \in \left( 0,\widetilde{\delta }\left( \lambda \right)
\right) .$ Next, using the first line of (\ref{9.12}), (\ref{9.40}) and
triangle inequality, we obtain \emph{\ }%
\begin{equation}
\left. 
\begin{array}{c}
\left\Vert V_{n}-V^{\ast }\right\Vert _{\left( H^{2}\left( \Omega
_{h,T}\right) \right) ^{3}}\leq \left( 1+\exp \left( \lambda \left(
b^{2}-a^{2}\right) \right) \right) \delta + \\ 
+C_{8}\theta ^{n}\left\Vert V_{\lambda ,\min }-V_{0}\right\Vert _{\left(
H^{2}\left( \Omega _{h,T}\right) \right) ^{3}},\text{ }\forall \delta \in
\left( 0,\widetilde{\delta }\left( \lambda \right) \right) ,\text{ }n=1,...%
\end{array}%
\right.  \label{9.41}
\end{equation}%
The similar estimate for $\left\Vert \beta _{n}-\beta ^{\ast }\right\Vert
_{L_{2}\left( \Omega \right) }+\left\Vert \gamma _{n}-\gamma ^{\ast
}\right\Vert _{L_{2}\left( \Omega \right) }$ follows immediately from (\ref%
{9.41}) and full analogs of formulas (\ref{8.17}). \ $\square $

\end{document}